%% file: main.tex
\documentclass[acmlarge]{acmart}

\usepackage{booktabs} 

\usepackage[ruled]{algorithm2e} 

\SetAlFnt{\small}
\SetAlCapFnt{\small}
\SetAlCapNameFnt{\small}
\SetAlCapHSkip{0pt}
\IncMargin{-\parindent}

\usepackage{graphicx, booktabs, 
}
\usepackage{xspace}
\usepackage{enumerate}
\usepackage{url}
\usepackage{verbatim}
\usepackage{multirow}
\usepackage{longtable}
\usepackage[textsize=scriptsize,textwidth=1cm]{todonotes}
\usepackage{setspace, array, rotating, caption, pdfpages, float}
\usepackage{pgfgantt}
\usepackage{lscape}
\usepackage{graphicx}
\usepackage{caption,subcaption}

 \usepackage{booktabs}
\usepackage{multirow}
\usepackage{graphicx}

\usepackage[xcolor,outerbars]{changebar}

\acmVolume{}
\acmNumber{}
\acmArticle{}
\acmYear{}
\acmMonth{0}
\acmArticleSeq{11}

\setcopyright{usgovmixed}

\acmDOI{0000001.0000001}

\received{}
\received{}
\received[accepted]{}

\begin{document}

\title{Emotionalism within People-Oriented Software Design}
\author{Mohammadhossein Sherkat}
\author{Tim Miller}
\author{Antonette Mendoza}
\author{Rachel Burrows}
\orcid{1234-5678-9012-3456}
\affiliation{
  \institution{University of Melbourne}
  \department{School of Computing and Information Systems}
  \streetaddress{}
  \city{}
  \state{}
  \postcode{}
  \country{}}

\begin{abstract}
In designing most software applications, much effort is placed upon the functional goals, which make a software system useful. However, the failure to consider emotional goals, which make a software system pleasurable to use, can result to disappointment and system rejection even if utilitarian goals are well implemented. Although several studies have emphasized the importance of people's emotional goals in developing software, there is little advice on how to address these goals in software system development process. 

This paper proposes a theoretically-sound and practical method by combining the theories and techniques of software engineering, requirements engineering and decision making.
The outcome of this study is the \emph {Emotional Goal Systematic Analysis Technique} (EG-SAT), which facilitates the process of finding software system capabilities to address emotional goals in software design. EG-SAT is an easy to learn and easy to use technique that helps analysts to gain insights in how to address people's emotional goals.
To demonstrate the method in use, a two-part evaluation is conducted. First, EG-SAT is used to analyze the emotional goals of potential users of a mobile learning application that provides information about low carbon living for trades people and professionals in the building industry in Australia. The results of using EG-SAT in this case study are compared with a professionally-developed baseline. Second, we ran a semi-controlled experiment in which 12 participants were asked to apply EG-SAT and another technique on part of our case study. The outcomes show that EG-SAT helped participants to both analyse emotional goals and gain valuable insights about the functional and non-functional goals for addressing people's emotional goals.

\end{abstract}


%
%


%
%


\thanks{Authors' addresses: School of Computing and Information Systems, University of Melbourne, Australia.}

\maketitle
\input{intro}

\input{background}

\input{proposed-method}
\input{evaluation}

\input{discussion}

\input{conclusions}

\begin{footnotesize}
\bibliography{bib}
\bibliographystyle{abbrv}
\end{footnotesize}



\end{document}

%% file: intro.tex
\section{Introduction}
\label{sec:intro}
\begin{quote}
``\emph{Interaction with technology is now as much about what people feel as it is about what people do.}'' John McCarthy \cite[p.\ 9]{mccarthy2007technology}.
\end{quote}


Developing and managing requirements is one of the most challenging parts of software engineering \cite{chung2012non}. In most requirements engineering techniques, much effort is placed upon the functional and non-functional goals of a system. However, the failure to consider emotional goals of users can result disappointment and system rejection, even if functional and non-functional goals are well implemented \cite{thewvalue, miller2015emotion}. In many applications, what users would like to feel (\emph{emotional goals}) are just as or even more important than what a system should accomplish \cite {callele2006emotional, bahsoon2005using}. As a result, stakeholders are demanding software that is more than just functional. People do not perceive a set of individual features in isolation, but instead evaluate the entire experience, including at an emotional level \cite {petermann2013gestalt}.
Consequently, if a software system is unable to attract users and appeal to their emotional needs, a likely consequence is that it will not be adopted or will lead to user frustration \cite {van2001interactive, platt2007software, Dix:2003:HI:1203012, 1663532, shneiderman2016designing}.

It is therefore important to take into account people's goals that will create a desire to engage with the system, as opposed to the fear of not completing a particular work task. These goals may be related to social values or emotions that people wish to feel \cite {sutcliffe2010analysing, miller2012understanding}. Following Marshall \cite{marshall2018agent}, we term such desires as \emph{emotional goals}.

In software engineering, emotional goals have typically been neglected as people assumed using a software system is a rational decision-making process. However, studies show that emotional goals are essential elements of considerations for interacting with anything that is designed to perform a function \cite {tzvetanova2007emotional}. As a result, in designing and implementing a successful software system, software engineers must decide on the most effective combination of software features to address what people want and desire.

Miller et al.\ \cite{miller2015emotion} show that emotional goals are not the same as what is often termed \emph{quality goals} \cite{miller2015emotion} because emotional goals are about people's reaction to a system rather than a property of the system itself. Emotional goals such as the desire to feel `Part of a community' or feel `a sense of worth' are formed from an individual's reflective emotional assessment of a system.

Addressing emotional goals in design is difficult for several reasons. 
First, emotional goals are the subjective part of people's consciousness, rather than the property of a software system. Second, people may be aware of their utilitarian goals, but they often are  unaware about their behavioural goals. Third, people's emotional goals are unstructured and often have a high level of ambiguity \cite{callele2006emotional,mendoza2013role}. 
These characteristics cause several complexities in software design: (i) the subjectivity of emotional goals make them difficult to be elicited; (ii) even if people's emotional goals are elicited, there is no universal method of representing them in a way that is useful for software engineering; and (iii) incorporating emotional goals into design is particularly challenging as it is difficult to understand how individual features support specific emotions.

The need to have adequate support for emotional goals is profound in social applications, such as platforms for enhancing social interaction, social networking and public health software systems. In these types of software systems, potential users are varied, often unknown with different kinds of personality, culture, goals, characteristics, needs and desires. We define these types of software systems as \emph{\textbf{P}eople-\textbf{O}riented \textbf{S}oftware} {(POS)} systems. 
POS systems can be differentiated from other systems as: (i) people are often not obliged to use a software system, (ii) people do not generally have well-defined roles and responsibilities and, (iii) people have different cultural and social backgrounds, and may conflicting as the same event can have wildly different emotional impacts on different people. These characteristics place greater importance on ensuring the motivation to engage is advocated for throughout the design process.



Although some approaches like iterative development or after-development fixing up \cite {robertson2001requirements, goguen1993techniques} can be helpful for addressing emotional goals in design, they are not ideal as people's emotional goals need to be understood before designing a software system. It may cost more if not integrated into the design process \cite {patel2009story}. Existing studies propose goals \cite {dardenne1993goal, anton1996goal}, personas \cite{sim2015developing} and scenario techniques \cite {holbrook1990scenario} to measure stakeholders' emotional perceptions of requirements around emotional-related qualities. However, there is a lack of a systematic approach to integrate emotions fully within requirements engineering and map these goals to design and implementation of systems through the software engineering life cycle \cite{miller2015emotion}.



In our previous work \cite {paper2}, we defined the \emph{Emotional Attachment Framework} --- a series of techniques for capturing people's emotional goals from data. In this paper, we argue that without a proper method for analyzing people's emotional goals to software specifications, efforts may not be led to action. 
For this reason, a technique for analyzing emotional goals and converting them to traditional functional and non-functional goals that can be analyzed by existing software engineering methodologies is necessary for system prototyping, verification, validation, and final implementation.
In this process, a key question for considering people's emotional goals in system design activities is: \emph{``How can system analysts achieve a better perception regarding the capabilities required to address emotional goals in a systematic manner?''}


To address this question, this paper aims to propose a method by combining the theories and techniques of software engineering, requirements engineering and decision making, and incorporate emotional goals from the beginning of the software development life-cycle. The outcome of this study is a technique entitled Emotional Goal Systematic Analysis Technique (EG-SAT), which provides an approach to facilitate the process of finding software system capabilities to address emotional goals in software design. 

We evaluate our method in two parts.
First, using an industry case study --- a mobile learning application for the building industry sector. 
We recruited $16$ participants and asked them to reflect their needs when using a mobile application for learning purposes. We analyzed and modelled the emotional goals of the key
stakeholders using the Emotional Attachment Framework \cite{paper2}. From the resulting models, we used EG-SAT to design and build a digital prototype. Via a subjective assessment of the digital prototype by 22 domain experts and end-users, compared with an alternative digital prototype also developed for the project, indicates that EG-SAT helped us find appropriate functional and non-functional goals for addressing people's emotional goals.

Second, we recruited $12$ participants with software engineering background and asked them to complete a series of tasks and answer a series of questions about EG-SAT and any proposed alternative analysis techniques. We measured the time and accuracy of their responses and then asked a set of qualitative questions around their preferences between the techniques. We then asked domain experts to analyze the participants' results and evaluate their output.
The results show that the EG-SAT is more time efficient, easier to use, easier to learn and can lead to more complete, correct and consistent results than the participant's chosen techniques.

The following section discusses some main concepts, complexity of people's emotional goals and prior efforts in considering people's emotional goals in design. In Section~\ref{sec:proposedmethod}, we present our method for analyzing people's emotional goals and this proposed method will be evaluated in Sections~\ref{sec:evaluation}. The last two sections of this paper are dedicated to discussion and conclusion.

%% file: background.tex
\section{Literature review and related work}
\label{sec:background}
Even though the importance of people's emotional goals have been emphasized in designing software systems and previous research has highlighted the growing need to consider emotional goals \cite{bentley2002putting, miller2015emotion, lopez2014one, miller2012understanding, lopez2014modelling, browne2002improving}, there is a huge gap between saying these are important and actually providing systematic techniques for incorporating them into design. 

Over the past three decades, several software engineering methodologies have developed to assist system analysts to capture goals and derive system functionality that supports them with various levels of abstraction and rigor \cite {chung2012non, song2010non, cysneiros2001framework, greenspan1982capturing}.
Despite the maturity of existing software development approaches, it is widely argued that a major focus of these approaches is on functional and non-functional requirements, and overlooking key drivers of that engagement; i.e. people's values and emotions \cite {bentley2002putting, draper1999analysing, gogueny1994requirements, hassenzahl2001engineering, krumbholz2000implementing,miller2015emotion, proynova2011investigating}. 
As a result, designing system's specification for addressing people's emotional goals remains a challenge.

In this section we briefly review 
the efforts in considering emotional goals in software system design and development. At the end of this section, we introduce the Function Analysis System Technique, from which we borrow several concepts.

\subsection{Emotional Goals in Software Engineering Domain}

From the historical perspective, people's emotions has been a part of design implicitly long before discussing its necessity explicitly \cite {demir2008field}. Many of these methods have been applied to convert several types of emotional goals into design products, but they (i) are descriptive techniques which usually recommend general advise without a repetitive and concrete process model, (ii) have been designed for hard products  and  their main concern is product appearance and, (iii) are not formal enough to be used directly as an input of software engineering techniques. As a result the proposed techniques in product design domain may not provide a systematic process and method that can be used in software development purposes. 

Although emotional goals is not such a well-studied topic in software engineering, there are some studies 
to deal with this type of requirements. 
In the following, we will review these approaches to see whether these methods or approaches can be used for addressing the emotional goals in software system design or not. 


A number of previous studies have merely limited themselves to discuss the importance of considering soft goals - like emotional goals - and they have not gone beyond this preliminary stage. For instance, Sutcliffe \cite {sutcliffe2009designing, sutcliffe2010analysing} argued the role of people's feelings and users' value as `soft issues' that can have different effects on requirements in system development process.


Some of the past studies only outlined methods for understanding and recognizing such soft goals and did not mention how they can be used in designing a software system.
As an example, Beyer and Holtzblatt \cite{beyer1999contextual} suggested `Contextual Design' as a user centred approach for considering people's norms in software design. This technique is similar to ethnography which users daily work is observed to understand and capture their insights for building the system. Although this technique can be used for discovering people's emotional goals, it does not particularly suggest a way for considering these requirements in design. 

Friedman \cite{friedman2013value} in `Value Sensitive Design' (VSD) proposed a technique for capturing the users' moral values, what related to human welfare and justice such as accountability and freedom from bias, and including them into design. 
However, as Le Dantec et al., \cite{le2009values} argued, VSD only includes a known set of values and does not have a technique for eliciting the other users' values. Friedman in VSD also does not propose a tool for design a specific function for addressing a specific value in design process. 



By reviewing the literature, we can conclude that the majority of the previous studies have only focused on representing the soft goals. 
Yu \cite {eric2009social} by proposing the \emph{i*} modelling notation tried to model soft goals 
\cite{samavi2009strategic}. In Yu's studies \cite {eric2009social, samavi2009strategic} besides classic non-functional requirements such as `reliable' or `secure', some emotional requirements are also included soft goals, such as `trustworthy', `flexible', `minimal intrusion' or `normal lifestyle'. However, \emph{i*} (i) does not provide enough explanation of how the soft goals analysis should be carried out in detail, (ii) it does not advise how the soft goals should be identified and used in design process and, (iii) it also does not separate  quality goals from emotional goals which we argue is important. 

Miller at al., \cite{miller2015emotion} argue that emotional desires should be treated as first-class citizens in software engineering methodology. For capturing the desired feelings of stakeholders they proposed a notation of modelling emotional goals in agent-oriented modelling entitled `People Oriented Software Engineering' (POSE). They have compared the POSE with \emph{i*} and reported that participants
found POSE more efficient in terms of average time for modeling emotional requirements as well as eliciting the correct emotional requirements. However, they do not propose a method to take these emotional goals to a design solution.

The last category of previous studies has been devoted to approaches that evaluate effects of software systems on users' emotions.  For instance, Bianchi-Berthouze and Lisetti \cite{bianchi2002modeling} for considering the users' emotions proposed `Model Of User Emotions' (MOUE). In this model, by recognizing the users' emotion from their facial expressions by processing movement, the system interprets and reacts according to the users' emotions. MOUE by using motion sensor attempts to recognize and manage emotional expressions and creates a basis for translating emotional content to computational systems. 
As we discussed in Section \ref{sec:intro}, recognizing the users' emotion after designing a software application is not our aim as we believe people's emotional goals need to be understood before designing a software system and be considered as a part of system design.



This section has reviewed the current advice to consider emotional goals in product and software system design process. Whilst the importance of consider such requirements has been widely emphasized, the past studies, to the best of our knowledge, provide general approaches for using the emotional goals in software system design process so that there is an opportunity to be more specific in software engineering.  



As we saw in this section, the general advice for incorporating soft goals into system design by the existing techniques have some limitations in software engineering domain. 
First, they are predominately representation techniques and most of them only work at visualization level by proposing new notations. Accordingly, current techniques cannot express emotional goals as there are no formal representation available for visualizing them. 
Second, the current techniques are not equipped with techniques for converting soft goals into software specifications. As the previous studies show, the challenge of converting soft goals into design choices remains generally unsolved and the current techniques cannot be used for analyzing people's emotional goals and incorporating them into design parameters
 \cite{bode2011tracing, xu2006architectural}. 
Third, the main focus of the previous studies in considering the emotional goals is limited to developing emotion recognition methods for evaluating usability issues. As we reviewed, they don't directly focus on people's emotional goals in the early stage of system design process as they only focus on understand users' feelings and emotions about a software system and its interface and not the people's emotional goals for designing a system.  
Fourth, the output of these techniques also cannot be traced back and find how people's emotional goals have been addressed by the system specification or be integrated within the system.
Although these approaches facilitate the process of exploring people's emotional goals, they may not suggest a process for turning these emotional goals into concrete software requirements that can be implemented.

\subsection{Function Analysis System Technique}
\begin{figure}[] 
\centering
\includegraphics [scale=0.65]{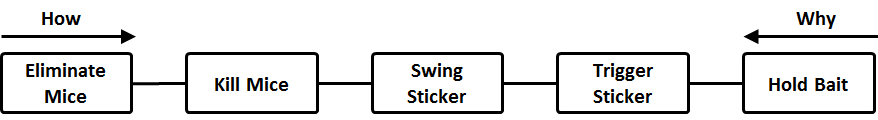}
\caption{ FAST Diagram for a Mouse Trap  \cite{valueanalysis}}
\label{fig:FASTSample}
\end{figure}

As the proposed technique in this paper was  built on Function Analysis System Technique (FAST) \cite{borza2011fast}, in this section we briefly review this technique. FAST is a tool for idea generation that was initially proposed by Charles W. Bytheway as a paper to the Society of American Value Engineers (SAVE) conference in 1965 and contributed significantly to perhaps the most important phase of the Value Methodology (VM) \cite {bytheway2007fast}. 
The success of using FAST in problem formulation and function analysis makes this technique an attractive candidate for middle level creativity and problem solving \cite{gerhardt2006managing, hanik2005ve}.
Figure \ref{fig:FASTSample} shows a simple sample FAST diagram for designing a mouse trap \cite{valueanalysis}. 

The main philosophy of FAST is that constructing a diagram in doing any analysis helps more than introspection analysis because diagramming rules organize the analysis process and structures its outputs. As a result, analytic efforts go forward in a structured manner and not fragmented as in a random process. Therefore, there is less likelihood of confusion and overlooking the different aspects of the subject under analysis. All this leads to a rigour and insightful analysis \cite{fowler1990value}. In addition, the determinate logic used in the FAST analysis and its diagram can be used for measuring the correctness and completeness of the analysis. The other advantage of the FAST is its higher capability to generate novel ideas. As the FAST provides systematic analysis method, it is expected to generate more ideas in the problem-solving process with higher quality than what expected from other problem-solving techniques such as brainstorming because it both provides constraints while giving freedom within those constraints \cite{kaufman2006stimulating}.
In this technique, there is no right or wrong model or result and analysts work until the real functionality of the system is identified, consensus is reached and analysts are satisfied that required requirements and functionality are expressed in the model. Using the FAST helps analysts to consider requirements as a complete unit, rather than analyzing them individually \cite{kaufman2006stimulating}. One of the advantages of FAST approach is its ability to represent function dependencies graphically. This ability facilitates stakeholders understanding and interpreting the system's functionality.

%% file: proposed-method.tex
\section{Proposed Method
}
\label{sec:proposedmethod}

System analysts need to produce the software requirements that will support emotional goals. Accordingly, any method for supporting emotional goals in software design needs to contain at least two elements: 1) a method for eliciting emotional goals; and 2) a method for analyzing emotional goals and proposing high-level solutions for addressing emotional goals. 
We address the first element using the Emotional Attachment Framework \cite{paper2}. In this paper, we address the second element by providing an analysis method that addresses emotional goals by deriving functional and quality goals to support them. This section describes the overall structure and the components of proposed method and its process model to facilitate the process of finding software system capabilities to address emotional goals in software design.


\subsection{Design Rationale}

In developing the proposed method in this paper, we have the following considerations and assumptions based on our discussions in the previous sections:

\begin{itemize}
\item 
People's emotional goals are usually addressed by functional and quality goals \cite{booch2005unified, miller2015emotion}. Accordingly, it is our assumption in this research that any method for analyzing emotional goals and addressing them needs to link emotional goals to functional and quality goals. 

\item
There are several mature analysis techniques for converting functional and quality goals into system specifications as we discussed in Section \ref{sec:background}. Accordingly, if we can find associated functional and quality goals for addressing emotional goals, existing software engineering methodologies can be used to develop these into a system. 


\item
Emotional goals are often high level and abstract \cite{paper2}. As a result, any proposed method for analyzing emotional goals should be able to break emotional goals down into more concrete concepts for analyzing them, and should be traceable.

\item Due to the unstructured and ambiguous nature of emotional goals \cite{paper2}, the analytic approaches that usually are used for defining the requirement specifications for functional and quality goals are not sufficient on their own. Finding functional and quality goals needs creative approaches that use idea-producing processes \cite{young2003technique} to generate a number of solutions. Once a list of potential requirement specifications is generated for addressing emotional goals, analytic processes can be used for selecting the feasible solution. 
\end{itemize}


Our approach aims to provide the right level of constraints to guide the process of analyzing emotional goals without overly constraining creativity and preserves the traceability of emotional goals through to the design features that support them.

\subsection{Emotional Attachment Framework}
Before analysing emotional goals, we first need to elicit them. One way to achieve this is via the Emotional Attachment Framework (EAF) from our previous work \cite{paper2}. Other requirements elicitation techniques may also be effective in uncovering emotional goals, however we give an overview of the EAF here to illustrate the range of emotional goals that are expected and possible to use as inputs to the analysis process.

\begin{figure}[] 
\small
\centering
\includegraphics [scale=0.50]{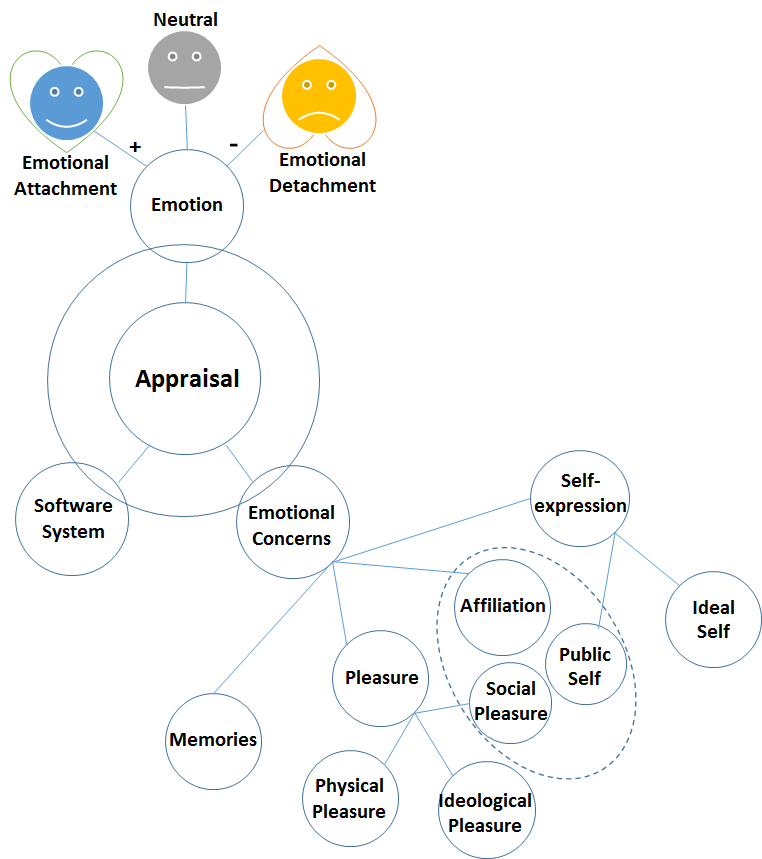}
\caption{Emotional Attachment Framework \cite{paper2}}
\label{fig:EAF}
\end{figure}

Emotional goals can be associated with a range of emotional attachment drivers such as ideal-self, public-self, social pleasure, etc (Figure~\ref{fig:EAF}). The hierarchical classification nature of EAF can reveal emotional goals that overlap, conflict or require consolidation in order to avoid confusion in requirement engineering process \cite{paper2}. It also provides analysts with an understanding of the underlying drivers of each emotional goal; such as whether the need is social, or not. For instance, a socially driven emotional goal may be better implemented in a software application with the ability to connect with other people. In short, a contextual understanding about the drivers of each emotional goal is valuable in the later consideration of software features that will support it. EAF categorizes the main drivers of forming emotional attachment under four categories including Self-expression, Affiliation, Pleasure and Memories \cite{paper2}. 

Elicited emotional goals provide a foundation for system design but do not necessarily provide complete insight required for the design process. In the current paper we present a method for analyzing emotional goals and converting them to something that existing software engineering techniques can use for system design purposes. 


\subsection{Emotional Goal Systematic Analysis Technique}
\label{EGDM}
In this section, we introduce our proposed method entitled \emph{Emotional Goal Systematic Analysis Technique (EG-SAT)}, which provides step-wise guidelines for analyzing people's emotional goals.  The EG-SAT enables the requirements engineering process to deal with the complexity of emotional goals analysis by using \emph{`How'} and \emph{`Why'} questions in the form of a structured diagram. The EG-SAT aims to help find functional and quality goals that address emotional goals.



We adopt the notation proposed by Sterling and Taveter \cite{sterling2009art}, shown in Figure \ref{fig:notattion}. The heart, cloud and parallelogram shapes represent the emotional, quality and functional goals respectively. These notations refer to the following definitions: 

\begin{figure}[] 
\small
\centering
\includegraphics [scale=0.62]{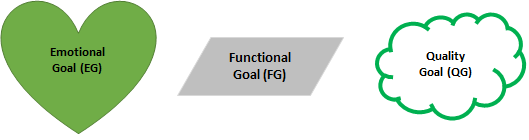}
\caption{\emph{EG-SAT} Notations}
\label{fig:notattion}
\end{figure}

\begin{itemize}
    \item \textbf{Functional Goal:} what people expect a software system should or should not do.
    \item \textbf{Quality Goal:} what people expect a software system should or should not be.
    \item \textbf{Emotional Goal:} what people would or would not expect to feel by using a software system.
\end{itemize}

Consider a social media application like Facebook\texttrademark. The goal of connecting friends is a functional goal that is quite different from the emotional goal of feeling connected. However, the functionality that connects friends  supports a feeling of connectedness.

\begin{figure}[]
\small
\centering
\includegraphics [scale=0.6]{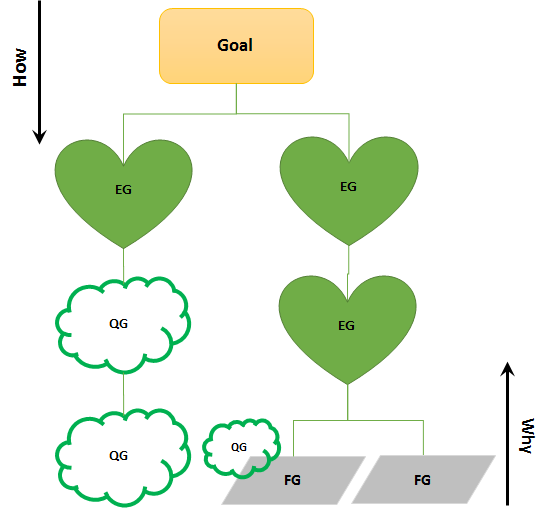}
\caption{ A Schematic View of EG-SAT Hierarchy} 
\label{fig:schematicEG-SAT}
\end{figure}

The EG-SAT promotes a hierarchical structure linking high-level emotional goals to more detailed emotional goals or functional and/or quality goals. In the EG-SAT hierarchy, any high-level emotional goal can be fulfilled by satisfying functional and/or quality goals in the next level of hierarchy. Figure \ref{fig:schematicEG-SAT} represent a schematic view of EG-SAT hierarchy. 
Each descending level in the EG-SAT hierarchy represents an increasingly-detailed description of the emotional, functional and quality goals. Functional and quality goals in the lower layers represent the high-level solutions that address emotional goals. 

The attached quality goal to the functional goal in Figure \ref{fig:schematicEG-SAT} shows that functional goals may be supported by quality goals. The \emph{`How'} and \emph{`Why'} arrows in Figure \ref{fig:schematicEG-SAT} --- as we will discuss in detail in Section~\ref{processmodel} --- show the direction of analysis of emotional goals in the EG-SAT. 
Each layer must contain all the goals (i.e. emotional, functional and quality) needed to ensure stakeholders achieve the next higher-level goals. 
This prevents extra goals that do not address emotional goals. Through the hierarchical structure of the EG-SAT, system analysts can always trace back to an emotional goal for specific functional and quality goals, meaning that the EG-SAT can be used for tracing the system capability in addressing people's emotional goals.

\subsection{Process Model}
\label{processmodel}

\begin{figure}[ht]
\small
\centering
\includegraphics [scale=0.6]{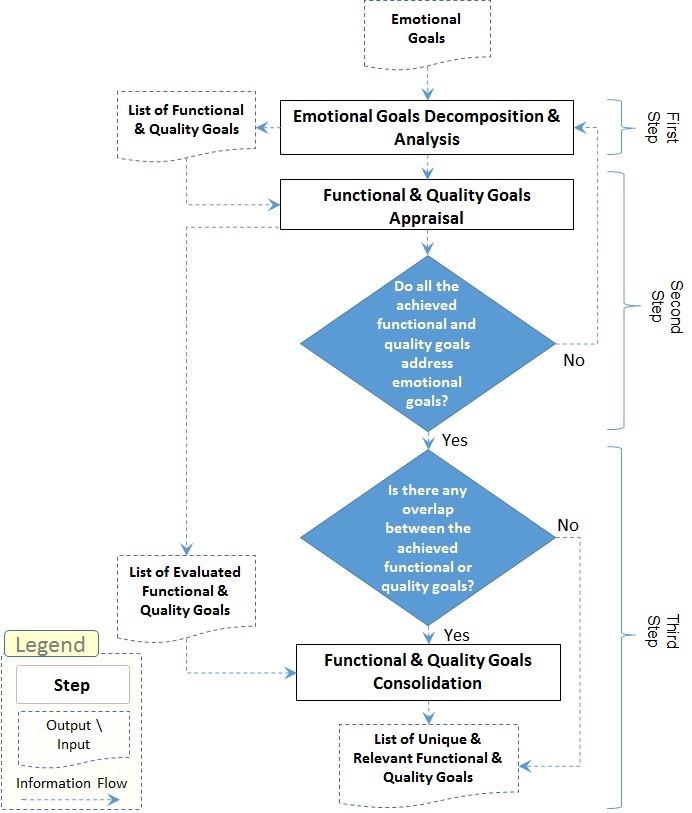}
\caption{Process Model} 
\label{fig:processmodel}
\end{figure}

In this section we outline a process model for the EG-SAT, outlined in Figure~\ref{fig:processmodel}. The input to the process model is a list of emotional goals, which can elicited using any technique such as EAF \cite {paper2}, POSE \cite{miller2015emotion}, ethnography \cite{neuman2005social}, etc.


The output of the EG-SAT is a list of functional and quality goals that support the emotional goals. These functional and quality goals will vary based on system analysts expertise and the project context. 
As shown in Figure \ref{fig:processmodel}, the EG-SAT process includes the following steps: (1) emotional goal decomposition and analysis; (2) functional and quality goal appraisal; and (3) functional and quality goal consolidation.


\subsubsection{Emotional Goal Decomposition and Analysis}
\label{FirstStep}


\medskip 

\begin{center}
\begin{tabular}{@{}llllp{5.5cm}@{}}
\toprule
\textbf{Task} & \textbf{Input} & \textbf{Technique} & \textbf{Output} & \textbf{Terminating Condition} \\ \midrule
Decomposition \&  & Emotional  & EG-SAT & List of Functional  & Achieve at least one relevant functional  
\\ 
Analysis & Goals & (\emph{`How'} question) & and Quality Goals & or quality goal for each emotional goal
\\
\bottomrule
\end{tabular}
\end{center}

\medskip

The first step aims to help system analysts to find sets of functional and quality goals for addressing the emotional goals. 
This starts by listing elicited emotional goals at the top of the EG-SAT hierarchy. The EG-SAT vertical hierarchy helps macro analysis of emotional goals until the key functional and quality goals for addressing emotional goals be identified. 
For this purpose, system analysts start asking the \emph{`How'} question that primes the analyst for getting down to a solution \cite{berger2014more}. This line of questioning and thinking is read from top to bottom. Asking this question in EG-SAT helps system analysts to get down to functional and quality goals that can be used for addressing emotional goals. 
A \emph{`How'} question may be answered by functional goals, quality goals, or even other emotional sub-goals.  It should be answered from the viewpoint of different stakeholders to capture their perspectives and create a variety of ideas.

\begin{figure}[]
    \small
    \centering
    \includegraphics [scale=0.7]{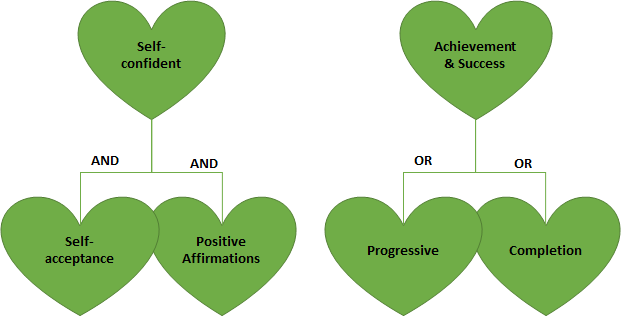}
    \caption{ Sample of  \emph{AND} and \emph{OR} decomposition} 
    \label{fig:andordec}
    \end{figure}

Emotional goals that are first elicited are usually high-level objectives. It means that if system analysts want to analyze abstract/combined emotional goals, they should first refine/decompose them to have a sufficient detail for further analysis. 
In the case of having an abstract/combined emotional goal, the \emph{`How'} question decomposes an emotional goal into a set of alternative emotional sub-goals such that satisfaction of one or all of them leads to the satisfaction of original emotional goal. Accordingly, there are two decomposition cases; (i) AND-decomposition: when every emotional sub-goal needs to be satisfied for the original emotional goal to be satisfied; and (ii) OR-decomposition: when the satisfaction of one emotional sub-goal is sufficient for the satisfaction of the original emotional goal. 
Figure \ref{fig:andordec} shows sample AND and OR decompositions. 
    
     For answering the \emph{`How'} question, different creative problem-solving techniques such as Brainstorming methods \cite{gallagher2013brainstorming}, the Systematic Inventive Technique \cite {goldenberg2001idea}, or the Theory of Inventive Problem Solving (TRIZ) \cite{altshuller1996and} can be used. In this step, it is important to avoid judgmental thinking as it constrains the initial creative process. What is important in asking the \emph{`How'} question is that possible answers should identify \emph{what} is to be designed and not \emph{how} it is to be implemented. In other words, the \emph{`How'} question determines how to fulfill the emotional goals with functional and quality goals, but not necessarily how these functional and quality goals should be exactly implemented in software engineering process. 
     
     Asking the \emph{`How'} question for each emotional goal should be continued until at least one functional or quality goal is achieved and system analysts are satisfied that a relevant functional or quality goal is identified.   In other words, the termination condition of asking the \emph{`How'} question is when there is no (sub-)emotional goals at the bottom of the EG-SAT hierarchy. This is because emotional goals are properties of people, not software, so cannot be implemented.
     
     
    
    \begin{figure}
  \begin{subfigure}[b]{0.43\textwidth}
    \includegraphics[width=\textwidth]{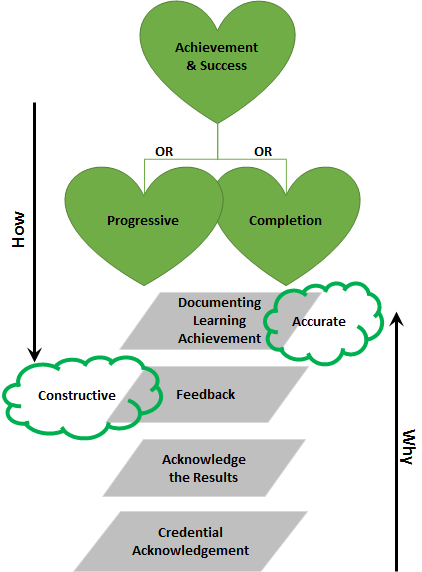}
    \caption{Sample of \emph{How} Question Analysis}
    \label{fig:egsatsample}
  \end{subfigure}
  \begin{subfigure}[b]{0.49\textwidth}
    \includegraphics[width=\textwidth]{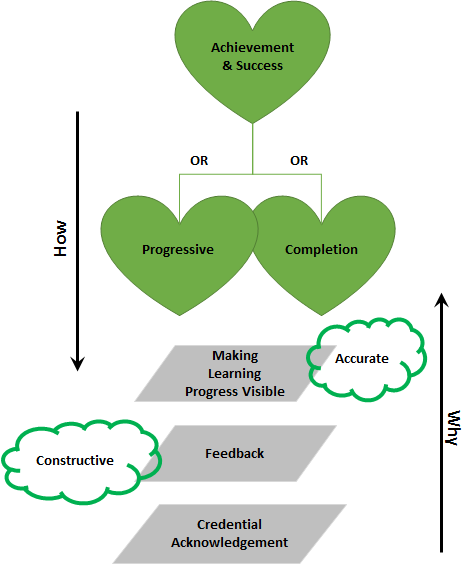}
    \caption{Sample of Merging Goals}
    \label{fig:merginggoals}
  \end{subfigure}
   \caption{Sample of Analysis and Merging}
   \label{fig:analysis-and-merging}
\end{figure}

 As an example, the possible answer for \emph{`How can sense of completion be addressed in a learning application?'} would be a functional goals like "Documenting Learning Achievement", "Credential Acknowledgement" and, "Giving Feedback". These functional goals can be implemented in different ways and through different software specifications and would require significant further analysis.
 Asking the \emph{`How'} question helps system analysts to avoid thinking just about a technical feature and miss the opportunity to engage in divergent thinking about other alternatives that can be used for addressing emotional goals.
 Figure \ref{fig:egsatsample} shows the EG-SAT hierarchy for a sample emotional goal. 
    
     
\subsubsection{Functional and Quality Goal Appraisal}
\label{secondstep}
\begin{center}

\medskip 

\label{SecondStepSummary}
\begin{tabular}{@{}lllll@{}}
\toprule
\textbf{Task} & \textbf{Input} & \textbf{Technique} & \textbf{Output} & \textbf {Terminating Condition} \\ \midrule
Appraisal & List of Functional  & EG-SAT & List of Evaluated Function  & Functional and Quality Goals   
\\ 
 & \& Quality Goals & (\emph{`Why'} question) &  and Quality Goals & are Relevant and Suitable
\\
\bottomrule
\end{tabular}
\end{center}

\medskip

The second step aims to help system analysts to evaluate the achieved functional and quality goals and make sure that they are relevant and address the emotional goals. 
%
In the first step, answering the \emph{`How'} questions, analysts should avoid judgmental thinking. In the second step, we answer \emph{`Why'} questions. This line of questioning is bottom-up, and  should switch from creative to critical thinking.
\emph{`Why'} questions are  interrogative questions whose primary goal is to help system analysts understand which functional and/or quality goals do not address the related emotional goal and should be eliminated or improved. This is an iterative process and can be repeated several times until the systems analyst is satisfied about the correctness of all of the achieved functional and quality goals. 
As an example, we use the EG-SAT hierarchy discussed in the previous step (Figure \ref{fig:egsatsample}). The possible answer to \emph{Why is "credential acknowledgement" necessary?} is ``because this functionality can help learners to see that they have completed learning tasks to a particular point". 


The \emph{`Why'} question can also be used for tracing functionality back to emotional goals. In other words, the \emph{`Why'} question can be used as a reverse engineering technique to determine which users' emotional goals can be addressed by existing system's specifications.

\subsubsection{Functional and Quality Goals Consolidation}


\medskip


\begin{center}
\label{thirdstepsummary}
\begin{tabular}{@{}lllll@{}}
\toprule
\textbf{Task} & \textbf{Input} & \textbf{Technique} & \textbf{Output} & \textbf{Terminating Condition} \\ 
\midrule
Consolidation & Evaluated Functional & Combination & List of Unique Functional & No Repetitive Goals  \\

& and Quality Goals & & and Quality Goals &  \\

\bottomrule
\end{tabular}
\end{center}

\medskip

The third step produces a list of non-repetitive functional and quality goals by asking whether there is any overlap between the proposed functional and quality goals. 

Consolidation is defined as an operation that combines two or more similar functional or quality goals. This occurs when two or more goals represent the same main concept. For this purpose, once the previous steps are complete, system analysts begin to group and consolidate similar functional and quality goals. As an example, \emph{"Documenting Learning Achievement"} and \emph{"Acknowledging the Results"} are two functional goals that have been achieved and evaluated in goal decomposition (Section \ref{FirstStep}) and goal appraisal (Section \ref{secondstep}) respectively. However, both of these functional goals refer to the same concept, \emph{"Making Learning Progress Visible"}. Accordingly, we can merge these two functional goals and replace them with a non-repetitive goal in the EG-SAT hierarchy (Figure \ref{fig:merginggoals}). Combined functional goals inherit their associated quality goals to the new functional goal resulting from the consolidation process.

\paragraph{Summary}
In EG-SAT, each emotional goal will be analyzed until it reaches to a specific functional and/or quality goal. The main focus of the proposed method is supporting the ideation process for addressing emotional goals. This method in this research is incremental and iterative so system analysts may switch between tasks as new ideas emerge. 
The lower level in EG-SAT shows functional and quality goals that the system analysts have more control over and can be used by system analysts for further analysis via existing software engineering methodologies.

%% file: evaluation.tex
\section{Evaluation}
\label{sec:evaluation}

Any methods in information systems and software engineering are developed to improve task performance in two ways: (i) improving the quality of the result; and (ii) reducing effort required to complete the task \cite{moody2003method, wieringa2014design, sonnenberg2011evaluation, prat2014artifact}. Accordingly, the aims of our evaluation in this research are as below:

\begin{itemize}
    \item Measuring the proposed method's effectiveness: to what extent can the proposed method help system analysts to understand the capabilities required to address emotional goals?
    \item Measuring the proposed method's efficiency: to what extent does the proposed method reduce the effort required to understand the capabilities required to address emotional goals?
\end{itemize}


For measuring the effectiveness of the proposed method, \emph{completeness}, \emph{correctness} and \emph{consistency} (3Cs) were chosen based on popular evaluation approaches in software engineering literature \cite{pennotti2009evaluating, zowghi2002three, lee2014software, pressman2005software}. In this paper, we define these quality measures as below in the context of our proposed method:

\begin{itemize}

\item \textbf{Completeness:} the proposed method leads to a complete analysis, 
if all the required functional and quality goals for addressing people's emotional goals have been specified.

\item \textbf{Correctness:} the proposed method leads to a correct analysis, 
if it represents the accurate and correct functional and quality goals for addressing people's emotional goals. 

\item \textbf{Consistency:}
According to the definition of consistency \cite{trochim2001research}, the proposed method leads to consistent results if (i) the achieved results are consistent within itself (internal consistency) and, (ii) the same result can be repeatedly derived (external consistency) \cite{leung2015validity}. 
Creative problem solving approaches such as EG-SAT are subjective, with different factors like context, experience and ideation technique used affecting the results. However, we believe that the high-level insights generated by the proposed method should be consistent and that different people will be able to consistently derive high-quality functional and quality goals that achieve the emotional goals. Thus, it is our hypothesis that if different people apply EG-SAT on the same data set, their main goals will be the same in concept.

\end {itemize} 


In this research by reviewing the literature \cite{moody2003method, davis1989perceived, fitzgerald1991validating, mendoza2010learnability, mendoza2010software}, the following metrics were used for measuring the method's efficiency: 

\begin{itemize}
    \item \textbf{Time:} this metric measures the time taken to complete the task by using a method.
    \item \textbf{Perceived Ease of Learning:} this metric measures to what extent a method would be easy to learn. 
    \item \textbf{Perceived Ease of Use:} this metric measures to what extent a method would be easy to use. 
    \item \textbf{Perceived Usefulness:} this metric measures to what extent a method would be effective in achieving its intended objectives.
    \item \textbf{Intention to Use:} this metric measures to what extent a person intends to use a particular method.
\end{itemize}

To address the evaluation aims and measuring the effectiveness and efficiency criteria, we undertook two main activities: (i) a case study analysis; and (ii) a semi-controlled experiment. 

We recruited 7 independent domain experts in our application domain. The case study is described in detail in Section \ref{sec:casestudy}. Five of the domain experts were exceedingly well versed in the domain of building sector and education in Australia and were heavily involved in sustainable urban design and energy efficiency. The other two domain experts were also part of the Cooperative Research Centre for Low-Carbon Living (CRCLCL) project and had large experience in designing and developing software applications. In each evaluation study domain experts worked independently, except for times when they required clarifications for doing the tasks. 

We also invited $17$ trainees and apprentices in the building and construction industry to participate in an evaluation. All of the participants were senior apprentices and trainees that, at the time of the study, had combined paid work and structured training between 36 to 48 months. $9$ out of $17$ trainees and participants had also three to five years of experience in industry/trade.
For doing the semi-controlled experiment, $12$ participants were recruited from a range of experience and expertise in software engineering domain and administered surveys to each. All the participants had a master's degree or higher in software engineering with three to five years working experience. Seven participants had specific training or experience in requirements engineering. Table \ref{sumofeval} shows a summary of evaluation techniques used in this study. 

In the following sections, 
we discuss the case study and semi-controlled experiment. For each activity, an overview of results will be discussed. The last part of this section is dedicated to reviewing some of the lessons that we learned. 



\input{./tables/sumofevaluation.tex}


\subsection{Case Study Analysis}
\label{sec:casestudy}

The case study is based on a real-world project, collaborating with industry partners, in order to develop an application to support carbon reduction and sustainable living. The target users of the application were those in the construction industry, with a key objective of the application being to increase motivation, enable collaboration and stimulate action in implementing low carbon living products and services. According to ``A Vision for Australia's Property and Construction Industry" \cite{hampson2004construction}, sustainable built environments and reducing greenhouse gas emissions is one of the top goals in the building sector in Australia over the next $20$ years.

 One of the authors of this paper applied EG-SAT in collaboration with development efforts within this project. The outcome was a mobile learning prototype application called \emph{Building Quality Passport} \cite{winfree2017learning}.
As part of this project, it was decided that the team should design and develop a mobile learning application entitled the \emph{Building Quality Passport} application to equip and motivate tradespeople and professionals in the building sector to engage with education on low-carbon building technologies and services. According to the anonymity, diversity and variety of \emph{Building Quality Passport}'s potential users, it is a people-oriented system.


\subsubsection{Method}
After eliciting emotional goals using the Emotional Attachment Framework \cite{paper2}, the first author led the application of EG-SAT to analyze the emotional goals of trainees and apprentices in the building sector in Australia for designing the digital prototype of the \emph {Building Quality Passport} that fulfilled the elicited emotional goals. Simultaneously, an application design and development company was contracted to undertake a design using the same elicited emotional goals, and applying their method of choice. Their choices were independent of the authors.

\paragraph{Emotional Goals} The first step was to find the list of emotional goals for the application to serve as input to the EG-SAT. For this, we applied the Emotional Attachment Framework \cite {paper2}. To gather sufficient data for the Emotional Attachment Framework, three different questionnaires were used with $16$ participants including $11$ building and trades trainees and apprentices, two employers and workplace mentors and, three training facilitators and trade teachers\footnote{The questionnaires are available at https://goo.gl/forms/qlkkQwN0L0SL3Ls73, https://goo.gl/forms/9EgIjz2CvZ8IIsZN2, https://goo.gl/forms/9EQmwcvM3QW91VSJ2 $~~$ and, https://goo.gl/forms/oh1jOxX64npgYLjb2}. The online questionnaires were focused on trainees and apprentices emotional goals for designing a mobile learning application entitled \emph{Building Quality Passport}. 

In each questionnaire, a series of general questions were asked, based on the following themes: 1) what should a mobile learning application do for you?; 2) how should it be?; and 3) how do you want to feel when using a mobile application for learning purposes? These questions were not asked directly as stated above, but were based on these themes.
For those participants with experience in using mobile learning applications, also some questions were asked regarding problems they experienced using mobile learning applications. The data was analyzed using the Emotional Attachment Framework approach \cite {paper2} to extract and model the key emotional goals and concerns by different stakeholders.
As a result, $56$ emotional goals were elicited and presented to the research group\footnote{The complete list is available at https://tinyurl.com/ya3fny4b}. Based on the similarity between the achieved emotional goals, the research group consolidated similar emotional goals to achieve a list of emotional goals. In this study the $56$ initial emotional goals were grouped into $24$ emotional goals\footnote {The complete list of emotional goals is available at https://tinyurl.com/y9c2hlqc}. Table \ref{emotionalgoals} represents a summary of results in the \emph{Building Quality Passport} case study. 


\begin{table}[]
\centering
\caption{Summary of Emotional Goals in \emph{Building Quality Passport} Case Study}
\label{emotionalgoals}
\footnotesize
\begin{tabular}{@{}lllll@{}}
\toprule
\textbf{ID} &\textbf{Emotional Goals} &\textbf{Sub-emotional Goals} & \textbf{Frequency} & \textbf{Emotional Attachment Driver} \\
\toprule
IP1-1 & Freedom and Flexibility & Sense of learning at my own pace & 4 & Ideological Pleasure \\
IP2 & Sense of time efficiency & $-$ & 4 & Ideological Pleasure \\
IP3 & Sense of trust in the information & $-$ & 4 & Ideological Pleasure \\
PP1 & Sense of reality & $-$ & 4 & Physical Pleasure \\
IS1 & Knowledgeable and Skillful  & $-$ & 3 & Ideal Self \\
PS1 & Professional & $-$ & 3 & Public Self \\
PS2 & Qualified & $-$ & 3 & Public Self \\
AF1 & Connected & $-$ & 3 & Affiliation \\
SP1 & Support and Assisted & $-$ & 3 & Social Pleasure \\
IS2-1 & Self-confident & Prepared & 2 & Ideal Self \\
IS2-2 & Self-confident & Sense of contribution & 2 & Ideal Self \\
IS3 & Sense of opportunity & $-$ & 2 & Ideal Self \\
IS4 & Sense of monetary (wealth) &  $-$ & 2 & Ideal Self \\
AF2 & Sense of networking & $-$ & 2 & Affiliation \\
IP4 & Sense of cost-effectiveness & $-$ & 2 & Ideological Pleasure \\
SP2 & Sense of promotion \& progression & $-$  & 2 & Social Pleasure \\
IS5 & Sense of being cutting edge & $-$  & 1 & Ideal Self \\
IS6-1 & Sense of achievement \& success & Progressive & 1 & Ideal Self \\
IS6-2 & Sense of achievement \& success & Sense of ongoing learning  & 1 & Ideal Self \\
IS6-3 & Sense of achievement \& success & Sense of growing strength & 1 & Ideal Self \\
IS6-4 & Sense of achievement \& success & Sense of completion & 1 & Ideal Self \\
IP1-2 & Freedom and Flexibility & In control & 1 & Ideological Pleasure \\
PP2 & Sense of fun & $-$ & 1 & Physical Pleasure \\
SP3 & Sense of competition & $-$ & 1 & Social Pleasure \\
\bottomrule

\end{tabular}
\end{table}

The fourth column in Table \ref{emotionalgoals} represents the frequency of each emotional goal; for example, four different emotional goals in the data had the same emotional concept of `learning at my own pace'. This indicates a higher importance and priority. By analyzing the on-line survey data we categorized the achieved emotional goals under the four main emotional drivers\footnote{Summary of results is available at https://tinyurl.com/y9l8mpdp}.

\paragraph{EG-SAT Application} We analyzed the set of emotional goals using our EG-SAT process model (Figure~\ref{fig:processmodel}).
The authors then conducted internal brain storming sessions to find some detailed design solutions (software features) for each functional and quality goal. Then the first author used these to design a digital prototype for the \emph{Building Quality Passport}\footnote{EG-SAT analysis is available at https://tinyurl.com/ybucrqwh and https://tinyurl.com/yahjsvuk respectively}. 

\paragraph{Baseline} To form a baseline for our method, the project funded the development of a second digital prototype. An application design and development company with several years of experience in this field and who has developed variety of successful projects was contracted to build this prototype to develop a digital prototype. As input to their process, they were given all previous findings, including data, the list of elicited emotional goals, and access to domain experts. A designer from this company gathered some additional data using its own data-gathering techniques and used their user experience methods for analyzing the emotional goals provided by the authors. Finally, they produced a digital prototype of their design. The authors of the present paper were not involved in this process except to provide the list of emotional goals and data.

\begin{figure}[ht]
\small
\centering
\includegraphics [scale=0.48]{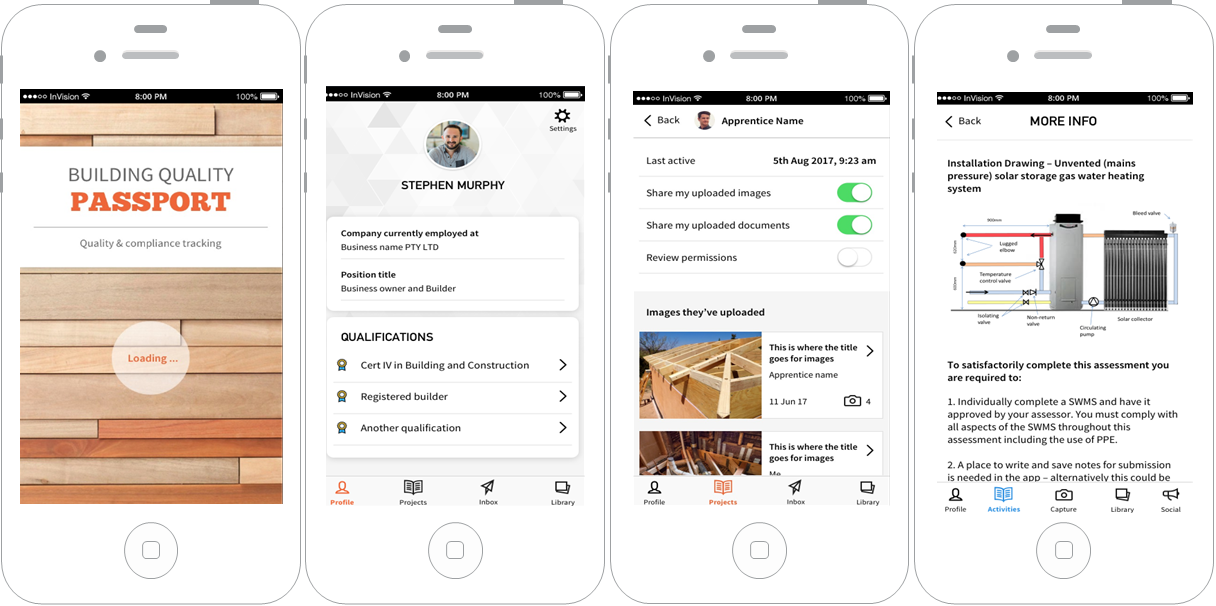}
\caption{Some Sample Screens of the Designed Digital Prototype - Baseline} 
\label{fig:versionA}
\end{figure}

\begin{figure}[ht]
\small
\centering
\includegraphics [scale=0.48]{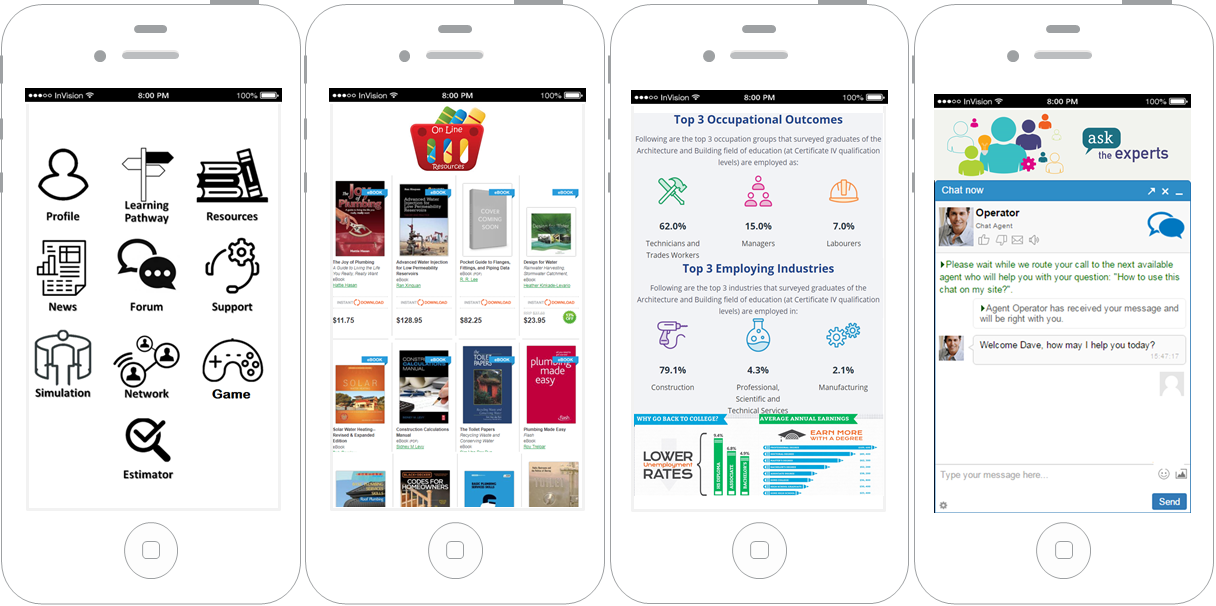}
\caption{Some Sample Screens of the Designed Digital Prototype - EG-SAT Version} 
\label{fig:versionB}
\end{figure}

\paragraph{Feature merging} Given that the scope of the EG-SAT ends with functional and quality goals, rather than design features, we merged common features (e.g., log-in and some basic features such as photo uploading and tagging) between both the baseline and the EG-SAT prototype. This enabled us to control for the variable of feature design, to ensure that participant ratings were about the features selected rather than how they were represented.
We call the professionally-developed prototype `Baseline' and ours the `EG-SAT Version'. Figure \ref{fig:versionA} and Figure \ref{fig:versionB} represent some sample screens of baseline and EG-SAT version respectively.

\subsubsection{Effectiveness Analysis}
\label{evalde}
We presented both prototypes to five of the domain experts (including three CRCLCL members and two training facilitator and trade teachers) and 17 trainees and apprentices over three focus groups, and gathered both qualitative feedback and quantitative ratings of the two prototypes. The following steps where conducted in the process of measuring the proposed method's effectiveness based on the case study:

\begin{itemize}

\item The list of elicited emotional goals used for developing the both versions of the digital prototypes (baseline and EG-SAT version) and the list of functional and quality goals (EG-SAT analysis output) used for developing the EG-SAT version were presented to the domain experts and trainees and apprentices separately and then they were asked to review them. 
 
\item Two versions of the digital prototypes (baseline and EG-SAT version) were presented to the participants. 
Note that these two versions were presented to  the domain experts and trainees and apprentices as two potential designs and both versions were products of the same joint project. The baseline version was presented by a non-author and the EG-SAT version by the first author of this paper, thus eliminating any potential subject bias.

\item The domain experts and trainees and apprentices were asked to complete two tasks: (i) compare the baseline version and EG-SAT version regarding their functionality and determine which version better addresses the elicited emotional goals; (ii) answer three open-ended questions:  (a) identify any additional functional or quality goals are needed for addressing the emotional goals; (b) identify any incorrect functional or quality goals are; and (c) identify any inconsistency within functional or quality goals. They were also asked to reflect their attitudes towards the designed digital prototypes, elicited emotional goals and associated functional and quality goals with other group members. 

For this purpose, the domain experts and trainees and apprentices were asked to answer a questionnaire\footnote{The questionnaire is available at https://tinyurl.com/yalts3wo}. 
One of the authors of the present paper took notes and recorded the vital points raised by the participates during the evaluation process.


 
 
\end{itemize}

To avoid any bias, both of the digital prototypes were developed by using the InVision\textsuperscript{\textregistered}\footnote{https://www.invisionapp.com/} platform to minimize the effects of interface and graphical design on the participants' judgment. 


\subsubsection{Results}
As we discussed at the start of Section \ref{sec:evaluation}, we investigated three evaluation criteria for measuring the effectiveness of the proposed method; completeness, correctness and consistency. In the following, a summary of results for effectiveness analysis will be discussed.  

\begin{figure}[h!]
  \begin{subfigure}[b]{0.49\textwidth}
    \includegraphics[width=\textwidth]{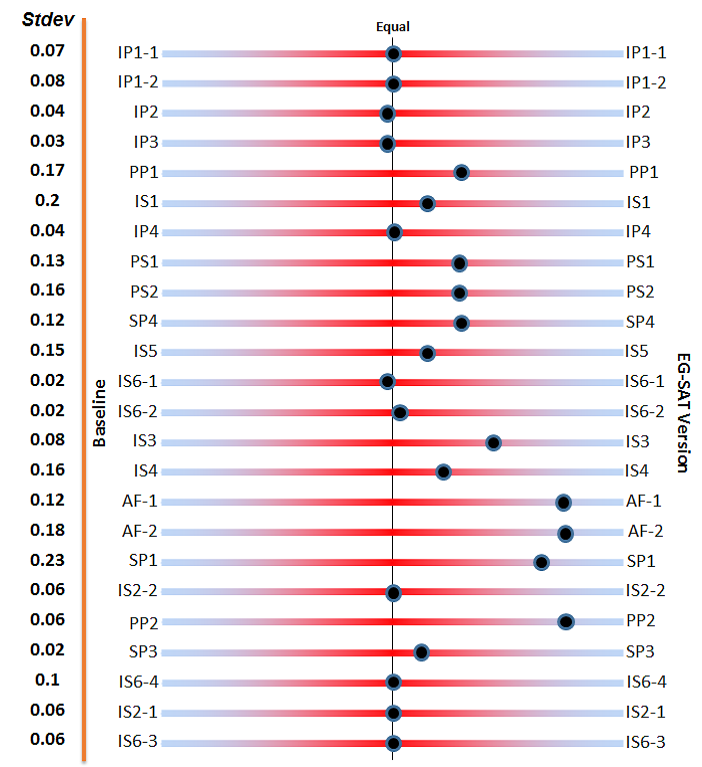}
    \caption{Domain Experts Analysis}
    \label{fig:resultsDE}
  \end{subfigure}
  \begin{subfigure}[b]{0.49\textwidth}
    \includegraphics[width=\textwidth]{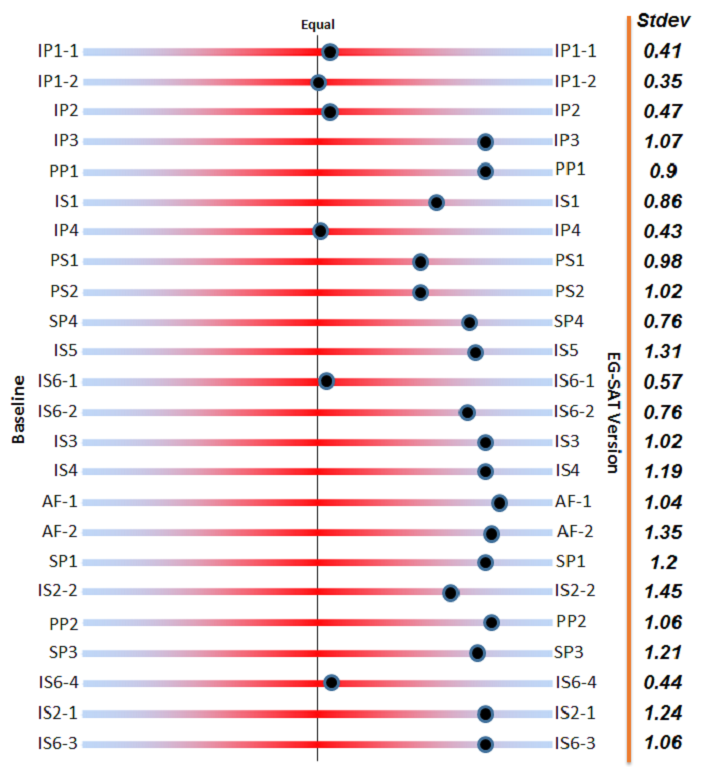}
    \caption{End-users Analysis}
    \label{fig:resultsEU}
  \end{subfigure}
   \caption{The Summary of Effectiveness Analysis Results }
\end{figure}

Figure \ref{fig:resultsDE} and Figure \ref{fig:resultsEU} show the mean and standard deviation of the responses by domain experts and end-users respectively as to which prototype best addresses the emotional goals.
The IDs on each side of distributions refer to each emotional goal as was explained in Table \ref{emotionalgoals}. The dark bullet shows the average of the participants' responses , therefore being closer to the left means the baseline is preferred, averaged over all participants in each study, and on the right, EG-SAT is preferred.


From these figures, we can see a strong preference over most emotional goals from both participant groups, which shows that the participants rated that the EG-SAT version better addresses the emotional goals than the baseline.
They also show that the average answers of the trainees and apprentices (end users) is higher than the domain experts. Understanding the main reason of this difference is not so complicated if we consider that the domain experts' main concern is learning while the trainees and apprentices are more interested in career outcomes. 

\paragraph{Completeness}
From our results, we conclude that EG-SAT, compared with the method used to derive the baseline, was more comprehensive for analyzing the emotional goals in terms of level of abstraction of emotional goals as well as the required solutions for addressing the emotional goals. 
The proposed method has a higher degree of completeness, both at the analytical level and at the level of presentation compared with the method used to derive the baseline.
Analyzing the domain experts responses to the qualitative questions shows that the domain experts  acknowledge the completeness and correctness of the EG-SAT version. 
For instance, domain experts acknowledged that considering the emotional goals in designing the application can engage users more. One domain expert stated:

\begin{quote}
\emph{``It [baseline] incorporates just some basic features. It was quiet basic in terms of current available. [The] EG-SAT version was more engaging,....it [EG-SAT version] has more modern feel.''}
\end{quote}



As the average answers of the domain experts show, EG-SAT version by address the potential users' emotional goals 
Another domain expert stated:

\begin{quote}
\emph{``[The EG-SAT version] is more fulfilling to use. .... it better responds to requirements.''}
\end{quote}

\paragraph{Correctness}
None of the domain experts raised a case that from their opinions is incorrect and needs to be changed or delete. It shows that from the domain experts' point of view the proposed method could lead to correct results. Responding to the  question about what they would  like add/change, one domain expert stated:

\begin{quote}
\emph{``Nothing, covers requirements very well''}
\end{quote}

One domain expert during presenting the EG-SAT version stated:

\begin{quote}
\emph{``It [EG-SAT version] makes it [learning process] a nicer, fuller experience that is not purely tidy-up in that one [baseline]."}
\end{quote}

\noindent and,

\begin{quote}
\emph{``They [trainees and apprentices] can use it [EG-SAT version] for healthy means, that is great!. We want it [EG-SAT version] to engage them [trainees and apprentices]."} 
\end{quote}


The trainees and apprentices comments in the questionnaire and in discussing others provided some valuable points.
For example, one stated:
     \begin{quote}
    \emph{``Somehow feel more comfortable and motivated. It stimulates me to explore it."}
    \end{quote}

Although in the trainees and apprentices comments there are some suggestions regarding the interface, none identified incorrect or incomplete requirements. It is an acknowledgment of completeness and correctness of the specified functional and quality goals in EG-SAT version for addressing the emotional goals.  

\paragraph{Consistency}

For measuring to what extent the achieved results are consistent within itself (internal consistency), we asked the domain experts and trainees and apprentices to determine whether any inconsistency exists (i) within the proposed quality goals; and (ii) within the proposed functional goals. No participant noted inconsistencies in the goals or prototype.




\subsection{Semi-controlled Experiment}

The goals of this semi-controlled experiment are two-fold. First, we evaluate whether our proposed method leads to complete, correct and consistent results when it is used by people independent of the authors (effectiveness). 
Second, we evaluate whether our proposed method leads to performance improvement in finding functional and quality goals (efficiency).

\subsubsection{Method}

To conduct the semi-controlled experiment, $12$ participants were recruited with a variety of experience and expertise in software engineering. Based on their responses in a screening survey, all of the participants had a master's degree or higher in the field of software engineering with three to five years working experience. Seven participants had specific  training in requirements engineering. 
The participants were asked applied both (i)EG-SAT and (ii) an alternative method of their choice to extract functional and quality goals from a subset of the emotional goals of the \emph{Building Passport Quality} case study, and measured the effectiveness and efficiency criteria (Table \ref{sumofeval}). 


The experiment used a within subject design, comparing two activities. In each activity, each participant was presented with the same sample of $12$ emotional goals of the \emph{Building Passport Quality} case study. To mitigate the potential bias, the sample emotional goals in each activity were selected randomly. 
In the first activity, participants were given the option to select any creative problem-solving techniques that they were familiar with to derive functional and quality goals from the 12 emotional goals. In the second activity, the EG-SAT technique was introduced to the participant, who was then asked to apply EG-SAT to derive functional and quality goals. To avoid an order effect, we counterbalanced by asking half of the participants to use EG-SAT in the first activity and half to use their method of choice in the second activity; and then switched. Participants in worked independently, and requested to only ask questions to clarify process.

\subsubsection{Efficiency and Effectiveness Analysis}

For measuring the efficiency and effectiveness criteria (Table \ref{sumofeval}), we gathered both qualitative feedback and quantitative ratings. As such, the following measures were taken:

\begin{itemize}

\item How long each participant used to complete his or her analysis in each round. There were no time limits on tasks and the average time spent by the participants was calculated in minutes.

\item Ease of learning, perceived ease of use and perceived usefulness. We administered a post-activity survey and asked participants in each group some open-ended qualitative questions about the used methods such as: (i) how comfortable they felt learning the method (easy to learn); (ii) how comfortable they would feel using the method for analyzing  emotional goals (easy to use); (iii) to what extent they found the used method useful (usefulness); and (iv) which method is their preference if they had to analyze people's emotional goals in the future? (intention to use)\footnote{The questionnaire  is available at https://tinyurl.com/ydaoj35p}. 
In the designed questionnaire some general questions were also asked about participants general feeling about the methods.

\end{itemize}

As it is difficult to objectively measure the participants' functional and quality goals, we went back to the domain experts to assess the participants' output. We supplied four domain experts (including two CRCLCL members and two others with software engineering background) with the sample emotional goals used by the participants in semi-controlled experiment and two lists of functional and quality goals by each participant, corresponding to the two rounds. Domain experts did not know which methods were used for each list. The domain experts were asked to do the following tasks: 

\begin{itemize}

\item Task 1: Review the lists and identify any inconsistency within the functional and quality goals suggested by the same participant in the two rounds. The output of this task was two lists of functional and quality goals which were deemed consistent from the domain experts point of view. 

\item Task 2: Review the lists and identify functional and quality goals that were deemed relevant for addressing the sample emotional goals. The output of this task was two lists of functional and quality goals which the domain experts endorsed their correctness. 

\item Task 3: Define unique functional and quality goals from what they had endorsed their correctness. For this purpose, the domain experts were asked to review the first list of the correct functional and quality goals and identify any functional and quality goals that they cannot map to any correct functional and quality goals suggested by the same participant in the second list of the correct functional and quality goals, and vice versa.

\end{itemize}




\subsubsection{Efficiency Results}

\begin{figure}[]
\small
\centering
\includegraphics [scale=0.53]{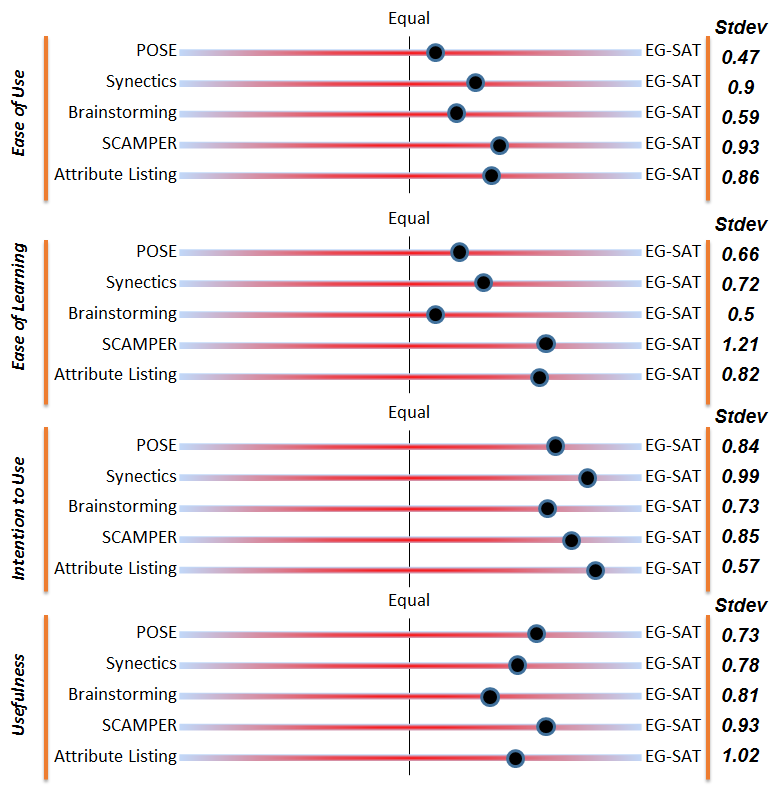}
\caption{The Summary of Efficiency Analysis Results} 
\label{fig:resultseff}
\end{figure}

Table \ref{semiexpresults} and Figure \ref{fig:resultseff} summarize the results for two experimental groups and evaluation metrics respectively. This shows the baseline methods uses and the average number of relevant proposed functional and quality goals by the participants, as determined by the domain experts.  

As Table \ref{semiexpresults} shows, over the $12$ participants, eight used brainstorming, and four others used techniques used Synectics, POSE, SCAMPER and Attribute Listing respectively. Figure \ref{fig:resultseff} shows the mean and standard deviation of the responses for the questions associated with qualitative metrics. From Figure \ref{fig:resultseff} we can see several expected results. First, participants have a stronger preference for using EG-SAT, largely because it has hierarchical layout, well-structured and easy to follow process model which lead to a more natural way to analyze emotional goals. Second, participants found the EG-SAT is easier to learn and use. 

For measuring to what extent the difference between the number of achieved functional and quality goals by using the baseline methods and EG-SAT and spent time are important, we conducted Wilcoxon signed-rank test \cite{sheskin2003handbook}. 
Our null hypothesis is that there is no statistically significant difference between the number of functional and quality goals achieved and time spent by baseline methods and EG-SAT. We conducted the Wilcoxon signed-rank test for at the $95\%$ level. Accordingly, the null hypothesis will be rejected if the \emph{p-value} be equal or less than $0.05$. 
Table \ref{semiexpresults} shows the p-values for number of functional and quality goals and time spent are $0.00222 (< 0.05)$, so are significant at this level. These results support our hypothesis that the EG-SAT can help produce higher quality functional and quality goals, and with less time/effort, compared to some other techniques.

\begin{table}[]
\caption{The Summary of semi-controlled experiment results }
\label{semiexpresults}
\resizebox{\textwidth}{!}{%
\begin{tabular}{lccccccc}
\toprule

& &\multicolumn{3}{c}{\textbf{Baseline Methods}} &\multicolumn{3}{c}{\textbf{EG-SAT Method}}  \\
\cmidrule(lr{1em}){3-5}
\cmidrule(lr{1em}){6-8}

\multicolumn{1}{l}{\multirow{4}{*}{\textbf{\begin{tabular}[c]{@{}c@{}}Baseline\\ Method\end{tabular}}}}& \multicolumn{1}{c}{\multirow{4}{*}{\textbf{\begin{tabular}[c]{@{}c@{}}Frequency\\ of use\end{tabular}}}} & \multicolumn{1}{c}{\multirow{4}{*}{\textbf{\begin{tabular}[c]{@{}c@{}}Ave. number of\\ relevant  FG*\end{tabular}}}} & \multicolumn{1}{c}{\multirow{4}{*}{\textbf{\begin{tabular}[c]{@{}c@{}}Ave. number of\\ relevant QG**\end{tabular}}}} &
\multicolumn{1}{c}{\multirow{4}{*}{\textbf{\begin{tabular}[c]{@{}c@{}}Time Spent \\ (minutes) \end{tabular}}}} &

\multicolumn{1}{c}{\multirow{4}{*}{\textbf{\begin{tabular}[c]{@{}c@{}}Ave. number of\\ relevant FG\\ \end{tabular}}}} & \multicolumn{1}{c}{\multirow{4}{*}{\textbf{\begin{tabular}[c]{@{}c@{}}Ave. number of \\ relevant QG\end{tabular}}}} & 

\multicolumn{1}{c}{\multirow{4}{*}{\textbf{\begin{tabular}[c]{@{}c@{}}Time Spent \\ (minutes)\end{tabular}}}} \\
\multicolumn{1}{c}{} & \multicolumn{1}{c}{} & \multicolumn{1}{c}{} & \multicolumn{1}{c}{} & \multicolumn{1}{c}{} & \multicolumn{1}{c}{} & \multicolumn{1}{c}{} & \multicolumn{1}{c}{} \\
\multicolumn{1}{c}{} & \multicolumn{1}{c}{} & \multicolumn{1}{c}{} & \multicolumn{1}{c}{} & \multicolumn{1}{c}{} & \multicolumn{1}{c}{} & \multicolumn{1}{c}{} & \multicolumn{1}{c}{} \\
\multicolumn{1}{c}{} & \multicolumn{1}{c}{} & \multicolumn{1}{c}{} & \multicolumn{1}{c}{} & \multicolumn{1}{c}{} & \multicolumn{1}{c}{} & \multicolumn{1}{c}{} & \multicolumn{1}{c}{} \\

\cmidrule(lr{1em}){1-5}
\cmidrule(lr{1em}){6-8}

POSE & 1 &6  &3  &16  &8  &4  & 10.8 \\
Synectics & 1 & 5 & 1 &15  &7  &2  & 11.5  \\
Brainstorming & 8  &7.25  & 2.875 &15.5  & 8.75 &4.625  & 11.4 \\
SCAMPER & 1 &5 & 0 & 12.5 & 6  & 3  & 9.2  \\
Attribute Listing & 1 & 5 & 0 & 14.5 & 8 & 2 & 13.5 \\
\cmidrule(lr{1em}){1-5}
\cmidrule(lr{1em}){6-8}
\multicolumn{2}{c}{\textbf{Ave.}} & \textbf{5.64} & \textbf{1.375} & \textbf{14.7} & \textbf{7.55} & \textbf{3.125} & \textbf{11.28} \\
\midrule
\multicolumn{3}{l}{
p-value for number of the FG: 0.00222} & \multicolumn{2}{l}{
p-value for number of the QG: 0.00222} & \multicolumn{3}{l}{
p-value for time spent: 0.00222}
\\ 

\bottomrule
\multicolumn{8}{l}{*FG: Functional Goal, **QG: Quality Goal}
\end{tabular}%
}
\end{table}

The participants' responses to the quantitative questions provide further evidence of the EG-SAT efficiency. Participants' responses support this argument that the EG-SAT is more useful method than other used techniques for analyzing the emotional goals in system analysis process. As Figure \ref{fig:resultseff} shows, all participants believe that the EG-SAT is easy to learn and use for analyzing people's emotional goals. For example, the following quotes are from three participants:

\begin{quote}
\emph{`` It's simplicity makes it both easy to learn and easy to apply."}
\end{quote}
    



\begin{quote}
\emph{``The method B [EG-SAT] is easy for me to use, it directs my thinking in a systematic manner, it is good recommended technique to analyze requirements prior to starting system development."}
\end{quote}

\begin{quote}
\emph{``it provides the relation between the emotional goals and associated functional and quality goals explicitly. It makes design justification and validation more straightforward."}
\end{quote}


However, one of the participants expressed preference for another technique. This participant commented that he/she preferred \emph{Brainstorming} because he/she believes:

\begin{quote}
    \emph{``It is more convenient for me. I used it several times before. I feel more confident with it, so I will go for it."} \ldots     \emph{``I think B [EG-SAT] is more structured so probably more useful for majority of people. It is also very useful for people how are not good at analysis as provides something to work with."}
    \end{quote}
    



\subsubsection{Effectiveness Results}

The results of domain experts analysis show that our proposed method helped participants. Table \ref{domainexperteval} shows the results of the domain experts' analysis. 

\paragraph{Completeness and Correctness}
As Table \ref{domainexperteval} shows, from the domain experts point of view, participants found more unique and relevant ways to address the emotional goals using the EG-SAT compared with other techniques. 
Although we acknowledge that achieving more unique and relevant functional and quality goals is not a benchmark for measuring the proposed method completeness and correctness, the uniqueness and relevance of achieved functional and quality goals support the claim that the EG-SAT has succeeded to provide the middle level creativity for analyzing people's emotional goals and higher capability to generate novel correct and complete ideas for addressing the emotional goals.



\input{./tables/domainexperteval.tex}

We conducted Wilcoxon signed-rank test at a 95\% level for these results, with the null hypothesis that there are no statistically significant differences between the number of unique and relevant functional/quality goals by using the EG-SAT and the other baseline techniques. Table \ref{domainexperteval} shows the results for, with p-values $0.00328$ and $0.00222$ respectively, therefore we reject the null hypotheses.

\paragraph{Internal Consistency:}
Although all the participants used the same data set and the main findings in the semi-controlled experiment were adapted from well-defined method (i.e. EG-SAT), there is a potential for inconsistency within the functional and quality goals suggested by each participant (internal consistency) and between the functional and quality goals suggested by the participants (external consistency).

As Table \ref{domainexperteval} shows that on average, participants' goals were more consistent when using EG-SAT than their baseline technique. The results show that the average number of inconsistent functional and  quality goals was more than double in the baseline methods. We again a Wilcoxon signed-rank test for a $95\%$ level, with the null hypothesis that there is no statistically significant difference between the number of inconsistent goals by using the EG-SAT and the other baseline techniques.
Table \ref{domainexperteval} shows the p-value for functional and quality goals are $0.00338$ and $0.00758$ respectively, therefore, we reject the null hypothesis.

\paragraph{External Consistency}
For measuring external consistency (consistency between participants), we used Cohen's Kappa values for measuring the inter-rater reliability to understand to what extent the proposed method lead to consistent results. Cohen's Kappa statistical measurements range from $-1.0$ to $1.0$; larger numbers represent better reliability and smaller numbers near zero suggest agreement has happened by chance \cite{saeed2013computer}.
As we discussed earlier, for such a creative problem solving approach as EG-SAT, the output is subjective and different factors like context, experience and the ideation technique used  can affect the results. However, it is our hypothesis that if different people use the EG-SAT for the same data set, regardless of the term used to describe the functional and quality goals, the main idea will be the same.


As a part of the process, one of the authors of the present paper analyzed the participants' responses and categorized the associated functional and quality goals for each emotional goal based on their similarities. Then the Cohen's Kappa values for measuring the proposed method consistency was calculated.
Table \ref{kappa} shows the statistical values for the Cohen's Kappa index. As all the Cohen's Kappa values for the both functional and quality goals are above 70\%, which is considered the minimum value for inter-rater agreement to be considered consistent \cite {boudreau2001validation}. It supports our hypothesis that the proposed method can lead to consistent results. 

\input{./tables/cohnkappa.tex}






\subsection{Limitations}

In the current paper, we considered only one case study and two different design projects. This is the trade-off of grounding this evaluation in a real industrial case study. Accordingly, generalizing the effectiveness and efficiency of the proposed method is limited and further case studies and evaluations with larger numbers would be a logical next step for this research. 
The case study is subject to bias, in that the participants in the semi-controlled experiment knew that the initial baseline results are used for comparison, which may have distorted their answers and lead to inaccuracies in our results. 
Additional studies would improve the validity of the proposed method for analyzing emotional goals.



%% file: tables/sumofevaluation.tex
\begin{table}[]
\centering
\caption{The Summary of Evaluation Techniques}
\label{sumofeval}
\begin{tabular}{llllc}
\toprule

\textbf{Evaluation Goal} & \textbf{Evaluation Method} & \textbf{Evaluation Criteria}  & \textbf{Coverage} & \multicolumn{1}{l}{\textbf{No.}} \\ \hline
\multirow{3}{*}{\textbf{Effectiveness}} & Case Study \& & Completeness, & Domain experts  & 7  \\ 
& Semi-controlled & Correctness \& & Educators \& apprentices  & 17 \\ 
& Experiment & Consistency  & Software engineers & 12  \\ \hline
\multirow{5}{*}{\textbf{Efficiency}} & & Time  &  \multirow{5}{*}{Software engineers} & \multirow{5}{*}{12} \\ 
& Semi- & Perceived Ease of Learning & &  \\ 
& controlled & Perceived Ease of Use & &  \\ 
& Experiment & Perceived Usefulness & &  \\ 
& & Intention to Use & &  \\ 
 \bottomrule
\end{tabular}%
\end{table}

%% file: tables/domainexperteval.tex
\begin{table}[h!]
\centering
\caption{The summary of domain experts evaluation results}
\label{domainexperteval}

\small

\begin{tabular}{lcccccccccc}
\toprule

\multicolumn{1}{l}{\multirow{3}{*}{\textbf{P*}}}
&\multicolumn{4}{c}{\textbf{Ave. \# of unique  \& relevant}} &\multicolumn{4}{c}{\textbf{Ave. \# of  inconsistent}}  \\
\cmidrule(lr{1em}){2-5}
\cmidrule(lr{1em}){6-9}
&\multicolumn{2}{c}{\textbf{Baseline }} &\multicolumn{2}{c}{\textbf{EG-SAT}} &\multicolumn{2}{c}{\textbf{Baseline }} &\multicolumn{2}{c}{\textbf{EG-SAT}}  \\
\cmidrule(lr{1em}){2-3}
\cmidrule(lr{1em}){4-5}
\cmidrule(lr{1em}){6-7}
\cmidrule(lr{1em}){8-9}

& {FG**} & {QG***} &{FG} &{QG}&{FG} & {QG}&{FG} &{QG}

\\







\hline
P1 & 1 & 0.25   & 1.25 & 2.75     & 1 & 1.75    & 0.25 & 0.5          \\
P2 & 0.25 & 0  & 3 & 2.25      & 1.25 & 2    & 0.75 & 0.5            \\
P3 & 0 & 0     & 2.25 & 2      & 2 & 2     & 0.5 & 1             \\
P4 & 0 & 0    & 3.25 & 3     & 1.25 & 1.75      & 0.5 & 0.75        \\
P5 & 1 & 0.75  & 2.25 & 3.25  & 1 & 1.5       & 0.75 & 0.5          \\
P6 & 0.25 & 1  & 3 & 2         & 1 & 1.5      & 0.5 & 0.25          \\
P7 & 0 & 0   & 4 & 0.75       & 1.5 & 1.5    & 0.75 & 0.5           \\
P8 & 0.5 & 0   & 0.25 & 3.25   & 1.5 & 1.75     & 0.25 & 0.75        \\
P9 & 1 & 0    & 2.25 & 3.5     & 1.5 & 1.5   & 0.75 & 0.75           \\
P10 & 0 & 0.25  & 3.5 & 3.75    & 0.5 & 1    & 0.5 & 0.25             \\
P11 & 0.75 & 0 & 3.75 & 2      & 1.25 & 0     & 0.25 & 0.75           \\
P12 & 0 & 0    & 2.25 & 1.25   & 1.5 & 0      & 0.25 & 0.75           \\

\cmidrule(lr){2-5} 
\cmidrule(lr){6-9}
Ave. & 0.4 & 0.19   & 2.58 & 2.48   &  1.3  & 1.4    & 0.5 & 0.6           \\
\bottomrule
\multicolumn{9}{l}{p-value for \# of unique \& relevant functional goals = 0.00328} \\  \multicolumn{9}{l}{p-value for \# of inconsistent functional goals = 0.00338 }\\ 
\multicolumn{9}{l}{p-value for \# of unique \& relevant quality goals = 0.00222}\\  \multicolumn{9}{l}{p-value for \# of inconsistent quality goals = 0.00758 }\\ 
\bottomrule
\multicolumn{9}{l} {*P: Participant ~~ **FG: Functional Goal~~ ***QG: Quality Goal}
\end{tabular}

\end{table}

%% file: tables/cohnkappa.tex
\begin{table}
\centering
\caption{Consistency Analysis - Cohen's Kappa Values }
\label{kappa}

\footnotesize

\resizebox{\textwidth}{!}{%
\begin{tabular}{c|l|llllllllllll}
\toprule
\multicolumn{2}{l}{\multirow{2}{*}{}} & \multicolumn{12}{c}{\textbf{Functional Goals}} \\ \cline{3-14} 
\multicolumn{2}{l}{} & P1 & P2 & P3 & P4 & P5 & P6 & P7 & P8 & P9 & P10 & P11 & P12 \\ \cline{3-14} 
\multirow{12}{*}{\textbf{\rotatebox[origin=c]{90}{Quality Goals}}}  
& P1 & N/A & 70.20 & 79.40 & 70.40 & 70.60 & 70.40 & 72.40 & 72.20 & 71.10 & 73.50 & 75.60 & 76.30 \\
 & P2 & 73.30 & N/A & 74.90 & 76.50 & 79.20 & 73.60 & 78.20 & 78.70 & 72.20 & 78.00 & 77.40 & 70.90 \\
 & P3 & 78.70 & 76.60 & N/A & 70.30 & 71.40 & 76.10 & 74.60 & 76.80 & 71.80 & 76.50 & 77.40 & 75.70 \\
 & P4 & 79.30 & 77.70 & 72.20 & N/A & 77.30 & 76.20 & 73.00 & 70.60 & 74.50 & 74.20 & 72.60 & 77.20 \\
 & P5 & 72.80 & 72.00 & 74.60 & 74.20 & N/A & 72.50 & 73.60 & 72.20 & 79.50 & 78.20 & 74.80 & 75.50 \\
 & P6 & 78.00 & 71.30 & 79.20 & 70.80 & 77.30 & N/A & 76.90 & 77.20 & 76.50 & 79.10 & 77.80 & 72.40 \\
 & P7 & 73.20 & 76.70 & 78.30 & 72.10 & 79.10 & 76.30 & N/A & 70.50 & 75.20 & 79.60 & 76.00 & 72.50 \\
 & P8 & 74.60 & 79.70 & 71.10 & 77.00 & 72.30 & 71.50 & 73.30 & N/A & 74.90 & 71.80 & 74.30 & 74.20 \\
 & P9 & 77.80 & 71.60 & 79.30 & 76.20 & 72.50 & 73.10 & 71.90 & 75.70 & N/A & 74.10 & 78.60 & 71.50 \\
 & P10 & 79.10 & 75.90 & 74.60 & 71.00 & 79.20 & 76.20 & 76.70 & 73.50 & 74.70 & N/A & 72.70 & 76.30 \\
 & P11 & 72.20 & 77.30 & 74.30 & 76.00 & 76.70 & 71.80 & 76.70 & 76.00 & 73.70 & 77.10 & N/A & 74.40 \\
 & P12 & 77.00 & 78.80 & 76.80 & 74.60 & 72.70 & 79.70 & 79.70 & 78.50 & 71.80 & 73.80 & 72.20 & N/A \\
\bottomrule
 \multicolumn{14}{l}{P*: Participant}

\end{tabular}%
}

\end{table}

%% file: discussion.tex
\section {Discussion and Lessons Learned}

In this paper, the research question, \emph{``How can system analysts achieve a better perception regarding the capabilities required to address emotional goals in a systematic manner?''} was addressed. 
To this effect, we designed and developed a new technique to facilitate the process of finding high-level software system capabilities in software design from emotional goals. 
Although emotional needs have received some attention in software engineering domain, to the best of our knowledge, the proposed method in the current paper is the first method that can help system analysts to systematically analyze people's emotional goals for system design purposes. 

The proposed method in this study was evaluated both via a semi-controlled experiment, and in designing a digital prototype of a mobile learning application. 
The results of case study analysis show that the EG-SAT can lead to improved outcomes of the designed digital prototype. The results of semi-controlled experiment also show that the EG-SAT is more useful method than other used techniques for analyzing the emotional goals in system analysis process as all the participants believe that the EG-SAT is easy to learn and use for analyzing people’s emotional goals. In terms of effectiveness metrics, the results show that participants found more unique, relevant and consistent ways to address the emotional goals using the EG-SAT compared with other techniques.
The results of the case study analysis and the semi-controlled experiment support the argument in this study that the EG-SAT can help system analysts to achieve a better perception regarding the capabilities required to address emotional goals.

As we experienced in this study, asking the \emph{How} question helped us to draw out the functional and quality goals while preventing us from thinking just about a technical solution. Using the EG-SAT questioning structure in analyzing and validating the emotional goals enabled greater focus, more freedom for creatively and intense exchange between the project team members. Asking the \emph{Why} question helped the research team to validate the results in a systematic manner and find the overlap between the suggested functional and quality goals. Further, questioning \emph{Why} helped us to avoid assumptions and logic traps and instead, trace the chain of causality in direct way from the proposed functional and quality goals to an emotional goal. 
However, we acknowledge the outcomes of questioning depends upon the knowledge and experience of the people involved.
    
As expected, the diagramming structure in the EG-SAT reduced confusion in analyzing the emotional goals. The EG-SAT hierarchy was effective at decomposing emotional goals. Based on our experience, the visualization of emotional, functional and quality goals in one diagram made research team conversions more advantageous compared to ordinary tables or text that we used in other projects\footnote{The diagram is available at 
https://tinyurl.com/yahjsvuk}. The EG-SAT hierarchy also helped us to represent which functional and quality goals must be fulfilled in the final application in order to an emotional goal to be fulfilled. 
    

%% file: conclusions.tex
\section{Conclusion and Future Work}

Over the past decade, we have seen a paradigm shift from designing software systems for satisfying functional goals towards the applications that are trying to enhance people's quality of life. Due to this, software engineers need to engage with emotional relationships between software systems and people. 
Although software engineering studies has highlighted the significant of emotional goals in software design, there has been little research to suggest a systematic and repetitive technique for this purpose. In this paper, we introduced a novel method for analyzing people's emotional goals and converting them to some functional and quality goals.

This study is significant from several aspects. First, it addresses the challenge and complexity of considering emotional goals in software system design process through investigating the emotional goals characteristics. Second, it proposes a novel method for analyzing people's emotional goals systematically in cooperation with functional and quality goals that allows in-depth analysis of emotional goals to build a software system, and provides a visual annotation for representing the analysis, which facilitates communication as well as documentation. Fourth, it provides traceability of emotional goals in system design through connecting the emotional goals to functional and quality goals. Finally, it bridges the gap between emotional goals elicitation and software system design process. 


In our future work, we will apply EG-SAT in other collaborative projects.
Our current study focused on evaluating the EG-SAT in analyzing people's emotional goals. However, as we discussed in this paper the EG-SAT can be used in reverse for analyzing the system maturity level in addressing people's emotional goals. As a part of our future study, we would like to understand this better. We encourage other researchers to use the proposed method in longer-term projects for the purpose of designing software systems.

\subsection*{Acknowledgements}
The authors would like to express their gratitude to participants for their participation in this study.  They also are thankful for the valuable feedback that they received on the proposed method and digital prototype from Tomi Winfree, Gavin Melles, Peter Graham,  Paul Goldacre,  and Tom Mansfield. This research is funded by the Australian Research Council Discovery Grant DP160104083 \emph{Catering for individuals' emotions in technology development}. The first author is funded by a University of Melbourne MIRS scholarship and a top-up from the CRC for Low-Carbon Living grant \emph{Increasing knowledge and motivating collaborative action on Low Carbon Living through team-based and game-based mobile learning}.

%% file: main.bbl
\begin{thebibliography}{10}

\bibitem{valueanalysis}
http://www.valueanalysis.ca/fast.php, [cited on-line: accessed 27/06/2018].
\newblock 2016.

\bibitem{altshuller1996and}
G.~Altshuller.
\newblock {\em And suddenly the inventor appeared: TRIZ, the theory of
  inventive problem solving}.
\newblock Technical Innovation Center, Inc., 1996.

\bibitem{anton1996goal}
A.~I. Anton.
\newblock Goal-based requirements analysis.
\newblock In {\em Requirements Engineering, 1996., Proceedings of the Second
  International Conference on}, pages 136--144. IEEE, 1996.

\bibitem{bahsoon2005using}
R.~Bahsoon, W.~Emmerich, and J.~Macke.
\newblock Using real options to select stable middleware-induced software
  architectures.
\newblock {\em IEE Proceedings-Software}, 152(4):167--186, 2005.

\bibitem{bentley2002putting}
T.~Bentley, L.~Johnston, and K.~von Baggo.
\newblock Putting some emotion into requirements engineering.
\newblock In {\em Proceedings of the 7th Australian workshop on requirements
  engineering}, pages 227--244, 2002.

\bibitem{berger2014more}
W.~Berger.
\newblock {\em A more beautiful question: The power of inquiry to spark
  breakthrough ideas}.
\newblock Bloomsbury Publishing USA, 2014.

\bibitem{beyer1999contextual}
H.~Beyer and K.~Holtzblatt.
\newblock Contextual design.
\newblock {\em interactions}, 6(1):32--42, 1999.

\bibitem{bianchi2002modeling}
N.~Bianchi-Berthouze and C.~L. Lisetti.
\newblock Modeling multimodal expression of user's affective subjective
  experience.
\newblock {\em User Modeling and User-Adapted Interaction}, 12(1):49--84, 2002.

\bibitem{bode2011tracing}
S.~Bode and M.~Riebisch.
\newblock Tracing the implementation of non-functional requirements.
\newblock {\em Non-Functional Properties in Service-Oriented Architecture:
  Requirements, Models and Methods}, pages 1--23, 2011.

\bibitem{booch2005unified}
G.~Booch.
\newblock {\em The unified modeling language user guide}.
\newblock Pearson Education India, 2005.

\bibitem{borza2011fast}
J.~Borza.
\newblock Fast diagrams: The foundation for creating effective function models.
\newblock {\em trizcon 2011}, 2011.

\bibitem{boudreau2001validation}
M.-C. Boudreau, D.~Gefen, and D.~W. Straub.
\newblock Validation in information systems research: a state-of-the-art
  assessment.
\newblock {\em MIS quarterly}, pages 1--16, 2001.

\bibitem{1663532}
J.~Brooks, F.P.
\newblock No silver bullet essence and accidents of software engineering.
\newblock {\em Computer}, 20(4):10--19, 1987.

\bibitem{browne2002improving}
G.~J. Browne and V.~Ramesh.
\newblock Improving information requirements determination: a cognitive
  perspective.
\newblock {\em Information \& Management}, 39(8):625--645, 2002.

\bibitem{bytheway2007fast}
C.~W. Bytheway.
\newblock {\em FAST creativity and innovation: Rapidly improving processes,
  product development and solving complex problems}.
\newblock J. Ross Publishing, 2007.

\bibitem{callele2006emotional}
D.~Callele, E.~Neufeld, and K.~Schneider.
\newblock Emotional requirements in video games.
\newblock In {\em Requirements Engineering, 14th IEEE International
  Conference}, pages 299--302. IEEE, 2006.

\bibitem{chung2012non}
L.~Chung, B.~A. Nixon, E.~Yu, and J.~Mylopoulos.
\newblock {\em Non-functional requirements in software engineering}, volume~5.
\newblock Springer Science \& Business Media, 2012.

\bibitem{cysneiros2001framework}
L.~M. Cysneiros, J.~C.~S. do~Prado~Leite, and J.~d. M.~S. Neto.
\newblock A framework for integrating non-functional requirements into
  conceptual models.
\newblock {\em Requirements Engineering}, 6(2):97--115, 2001.

\bibitem{dardenne1993goal}
A.~Dardenne, A.~Van~Lamsweerde, and S.~Fickas.
\newblock Goal-directed requirements acquisition.
\newblock {\em Science of computer programming}, 20(1):3--50, 1993.

\bibitem{davis1989perceived}
F.~D. Davis.
\newblock Perceived usefulness, perceived ease of use, and user acceptance of
  information technology.
\newblock {\em MIS quarterly}, pages 319--340, 1989.

\bibitem{demir2008field}
E.~Demir.
\newblock The field of design and emotion: Concepts, arguments, tools, and
  current issues.
\newblock {\em Journal of the Faculty of Architecture}, 25(1):135--152, 2008.

\bibitem{Dix:2003:HI:1203012}
A.~Dix, J.~E. Finlay, G.~D. Abowd, and R.~Beale.
\newblock {\em Human-Computer Interaction (3rd Edition)}.
\newblock Prentice-Hall, Inc., 2003.

\bibitem{draper1999analysing}
S.~W. Draper.
\newblock Analysing fun as a candidate software requirement.
\newblock {\em Personal Technologies}, 3(3):117--122, 1999.

\bibitem{eric2009social}
S.~Y. Eric.
\newblock Social modeling and i*.
\newblock In {\em Conceptual Modeling: Foundations and Applications}, pages
  99--121. Springer, 2009.

\bibitem{fitzgerald1991validating}
G.~Fitzgerald.
\newblock Validating new information systems techniques: a retrospective
  analysis.
\newblock {\em Information systems research: Contemporary approaches and
  emergent traditions}, pages 657--672, 1991.

\bibitem{fowler1990value}
T.~C. Fowler.
\newblock {\em Value analysis in design}.
\newblock CRC Press, 1990.

\bibitem{friedman2013value}
B.~Friedman, P.~H. Kahn~Jr, A.~Borning, and A.~Huldtgren.
\newblock Value sensitive design and information systems.
\newblock In {\em Early engagement and new technologies: Opening up the
  laboratory}, pages 55--95. Springer, 2013.

\bibitem{gallagher2013brainstorming}
S.~Gallagher.
\newblock {\em Brainstorming: Views and interviews on the mind}.
\newblock Andrews UK Limited, 2013.

\bibitem{gerhardt2006managing}
D.~J. Gerhardt and P.~I. Rand.
\newblock Managing value engineering in new product development.
\newblock {\em Value World}, 29(2):26, 2006.

\bibitem{goguen1993techniques}
J.~A. Goguen and C.~Linde.
\newblock Techniques for requirements elicitation.
\newblock {\em RE}, 93:152--164, 1993.

\bibitem{gogueny1994requirements}
J.~A. Gogueny.
\newblock Requirements engineering as the reconciliation of technical and
  social issues.
\newblock {\em Requirements Engineering: Social and Technical Issues, edited
  with Marina Jirotka, Academic Press}, pages 165--199, 1994.

\bibitem{goldenberg2001idea}
J.~Goldenberg, D.~R. Lehmann, and D.~Mazursky.
\newblock The idea itself and the circumstances of its emergence as predictors
  of new product success.
\newblock {\em Management science}, 47(1):69--84, 2001.

\bibitem{greenspan1982capturing}
S.~J. Greenspan, J.~Mylopoulos, and A.~Borgida.
\newblock Capturing more world knowledge in the requirements specification.
\newblock In {\em Proceedings of the 6th international conference on Software
  engineering}, pages 225--234. IEEE Computer Society Press, 1982.

\bibitem{hampson2004construction}
K.~D. Hampson and P.~Brandon.
\newblock {\em Construction 2020-A vision for Australia's property and
  construction industry}.
\newblock CRC Construction Innovation, 2004.

\bibitem{hanik2005ve}
P.~Hanik and J.~J. Kaufman.
\newblock Ve/triz: A technology partnership.
\newblock In {\em SAVE Conference}, 2005.

\bibitem{hassenzahl2001engineering}
M.~Hassenzahl, A.~Beu, and M.~Burmester.
\newblock Engineering joy.
\newblock {\em IEEE Software}, 18(1):70, 2001.

\bibitem{holbrook1990scenario}
H.~Holbrook~III.
\newblock A scenario-based methodology for conducting requirements elicitation.
\newblock {\em ACM SIGSOFT Software Engineering Notes}, 15(1):95--104, 1990.

\bibitem{kaufman2006stimulating}
J.~J. Kaufman and R.~Woodhead.
\newblock {\em Stimulating innovation in products and services: with function
  analysis and mapping}, volume~45.
\newblock John Wiley \& Sons, 2006.

\bibitem{krumbholz2000implementing}
M.~a. Krumbholz, J.~Galliers, N.~Coulianos, and N.~Maiden.
\newblock Implementing enterprise resource planning packages in different
  corporate and national cultures.
\newblock {\em Journal of Information Technology}, 15(4):267--279, 2000.

\bibitem{le2009values}
C.~A. Le~Dantec, E.~S. Poole, and S.~P. Wyche.
\newblock Values as lived experience: evolving value sensitive design in
  support of value discovery.
\newblock In {\em Proceedings of the SIGCHI conference on human factors in
  computing systems}, pages 1141--1150. ACM, 2009.

\bibitem{lee2014software}
M.-C. Lee.
\newblock Software quality factors and software quality metrics to enhance
  software quality assurance.
\newblock {\em British Journal of Applied Science \& Technology},
  4(21):3069--3095, 2014.

\bibitem{leung2015validity}
L.~Leung.
\newblock Validity, reliability, and generalizability in qualitative research.
\newblock {\em Journal of family medicine and primary care}, 4(3):324, 2015.

\bibitem{lopez2014one}
A.~A. Lopez-Lorca, T.~Miller, S.~Pedell, A.~Mendoza, A.~Keirnan, and
  L.~Sterling.
\newblock One size doesn't fit all: diversifying the user using personas and
  emotional scenarios.
\newblock In {\em Proceedings of the 6th International Workshop on Social
  Software Engineering}, pages 25--32. ACM, 2014.

\bibitem{lopez2014modelling}
A.~A. Lopez-Lorca, T.~Miller, S.~Pedell, L.~Sterling, and M.~K. Curumsing.
\newblock Modelling emotional requirements, 2014.

\bibitem{marshall2018agent}
J.~Marshall.
\newblock Agent-based modelling of emotional goals in digital media design
  projects.
\newblock In {\em Innovative Methods, User-Friendly Tools, Coding, and Design
  Approaches in People-Oriented Programming}, pages 262--284. IGI Global, 2018.

\bibitem{mccarthy2007technology}
J.~McCarthy and P.~Wright.
\newblock {\em Technology as experience}.
\newblock MIT press, 2007.

\bibitem{mendoza2010software}
A.~Mendoza, J.~Carroll, and L.~Stern.
\newblock Software appropriation over time: from adoption to stabilization and
  beyond.
\newblock {\em Australasian Journal of Information Systems}, 16(2), 2010.

\bibitem{mendoza2013role}
A.~Mendoza, T.~Miller, S.~Pedell, L.~Sterling, et~al.
\newblock The role of users' emotions and associated quality goals on
  appropriation of systems: two case studies.
\newblock In {\em 24th Australasian Conference on Information Systems}.
  Citeseer, 2013.

\bibitem{mendoza2010learnability}
A.~Mendoza, L.~Stern, and J.~Carroll.
\newblock Learnability’as a positive influence on technology use.
\newblock In {\em Electronic Proceedings of the 4th International
  Multi-Conference on Society, Cybernetics and Informatics, Retrieved from
  http://www. iiis. org/CDs2010/CD2010SCI/IMSCI\_2010/index. asp}, 2010.

\bibitem{miller2015emotion}
T.~Miller, S.~Pedell, A.~A. Lopez-Lorca, A.~Mendoza, L.~Sterling, and
  A.~Keirnan.
\newblock Emotion-led modelling for people-oriented requirements engineering:
  the case study of emergency systems.
\newblock {\em Journal of Systems and Software}, 105:54--71, 2015.

\bibitem{miller2012understanding}
T.~Miller, S.~Pedell, L.~Sterling, F.~Vetere, and S.~Howard.
\newblock Understanding socially oriented roles and goals through motivational
  modelling.
\newblock {\em Journal of Systems and Software}, 85(9):2160--2170, 2012.

\bibitem{moody2003method}
D.~L. Moody.
\newblock The method evaluation model: a theoretical model for validating
  information systems design methods.
\newblock {\em ECIS 2003 proceedings}, page~79, 2003.

\bibitem{neuman2005social}
W.~L. Neuman.
\newblock {\em Social research methods: Quantitative and qualitative
  approaches}, volume~13.
\newblock Allyn and Bacon Boston, 2005.

\bibitem{patel2009story}
C.~Patel and M.~Ramachandran.
\newblock Story card maturity model (smm): A process improvement framework for
  agile requirements engineering practices.
\newblock {\em JSW}, 4(5):422--435, 2009.

\bibitem{pennotti2009evaluating}
M.~Pennotti, R.~Turner, and F.~Shull.
\newblock Evaluating the effectiveness of systems and software engineering
  methods, processes and tools for use in defense programs.
\newblock In {\em Systems Conference, 2009 3rd Annual IEEE}, pages 319--322.
  IEEE, 2009.

\bibitem{petermann2013gestalt}
B.~Petermann.
\newblock {\em The Gestalt theory and the problem of configuration}.
\newblock Routledge, 2013.

\bibitem{platt2007software}
D.~S. Platt.
\newblock {\em Why Software Sucks--and what You Can Do about it}.
\newblock Addison-Wesley Professional, 2007.

\bibitem{prat2014artifact}
N.~Prat, I.~Comyn-Wattiau, and J.~Akoka.
\newblock Artifact evaluation in information systems design-science research-a
  holistic view.
\newblock In {\em PACIS}, page~23, 2014.

\bibitem{pressman2005software}
R.~S. Pressman.
\newblock {\em Software engineering: a practitioner's approach}.
\newblock Palgrave Macmillan, 2005.

\bibitem{proynova2011investigating}
R.~Proynova, B.~Paech, S.~H. Koch, A.~Wicht, and T.~Wetter.
\newblock Investigating the influence of personal values on requirements for
  health care information systems.
\newblock In {\em Proceedings of the 3rd Workshop on Software Engineering in
  Health Care}, pages 48--55. ACM, 2011.

\bibitem{robertson2001requirements}
S.~Robertson.
\newblock Requirements trawling: techniques for discovering requirements.
\newblock {\em International Journal of Human-Computer Studies},
  55(4):405--421, 2001.

\bibitem{saeed2013computer}
K.~Saeed, R.~Chaki, A.~Cortesi, and S.~Wierzcho{\'n}.
\newblock {\em Computer Information Systems and Industrial Management: 12th
  IFIP TC 8 International Conference, CISIM 2013, Krakow, Poland, September
  25-27, 2013, Proceedings}, volume 8104.
\newblock Springer, 2013.

\bibitem{samavi2009strategic}
R.~Samavi, E.~Yu, and T.~Topaloglou.
\newblock Strategic reasoning about business models: a conceptual modeling
  approach.
\newblock {\em Information Systems and e-Business Management}, 7(2):171--198,
  2009.

\bibitem{paper2}
M.~Sherkat, A.~Mendoza, T.~Miller, and R.~Burrows.
\newblock Emotional attachment framework for people-oriented software.
\newblock {\em arXiv preprint arXiv:1803.08171}, 2018.

\bibitem{sheskin2003handbook}
D.~J. Sheskin.
\newblock {\em Handbook of parametric and nonparametric statistical
  procedures}.
\newblock crc Press, 2003.

\bibitem{shneiderman2016designing}
B.~Shneiderman, C.~Plaisant, M.~S. Cohen, S.~Jacobs, N.~Elmqvist, and
  N.~Diakopoulos.
\newblock {\em Designing the user interface: strategies for effective
  human-computer interaction}.
\newblock Pearson, 2016.

\bibitem{sim2015developing}
W.~W. Sim and P.~Brouse.
\newblock Developing ontologies and persona to support and enhance requirements
  engineering activities--a case study.
\newblock {\em Procedia Computer Science}, 44:275--284, 2015.

\bibitem{song2010non}
X.~Song, Z.~Duan, and C.~Tian.
\newblock Non-functional requirements elicitation and incorporation into class
  diagrams.
\newblock {\em Intelligent Information Processing V}, pages 72--81, 2010.

\bibitem{sonnenberg2011evaluation}
C.~Sonnenberg and J.~Vom~Brocke.
\newblock Evaluation patterns for design science research artefacts.
\newblock In {\em European Design Science Symposium}, pages 71--83. Springer,
  2011.

\bibitem{sterling2009art}
L.~Sterling and K.~Taveter.
\newblock {\em The art of agent-oriented modeling}.
\newblock MIT Press, 2009.

\bibitem{sutcliffe2009designing}
A.~Sutcliffe.
\newblock Designing for user engagement: Aesthetic and attractive user
  interfaces.
\newblock {\em Synthesis lectures on human-centered informatics}, 2(1):1--55,
  2009.

\bibitem{sutcliffe2010analysing}
A.~Sutcliffe and S.~Thew.
\newblock Analysing "people" problems in requirements engineering.
\newblock In {\em 2010 ACM/IEEE 32nd International Conference on Software
  Engineering}, volume~2, pages 469--470. IEEE, 2010.

\bibitem{thewvalue}
S.~Thew and A.~Sutcliffe.
\newblock Value-based requirements engineering: method and experience.
\newblock {\em Requirements Engineering}, pages 1--22, 2017.

\bibitem{trochim2001research}
W.~M. Trochim and J.~P. Donnelly.
\newblock Research methods knowledge base.
\newblock 2001.

\bibitem{tzvetanova2007emotional}
S.~Tzvetanova, M.-X. Tang, and L.~Justice.
\newblock Emotional web usability evaluation.
\newblock {\em Human-Computer Interaction. HCI Applications and Services},
  pages 1039--1046, 2007.

\bibitem{van2001interactive}
M.~Van~Harmelen.
\newblock {\em Interactive system design using {OO} \& {HCI} methods}.
\newblock Addison-Wesley, 2001.

\bibitem{wieringa2014design}
R.~J. Wieringa.
\newblock {\em Design science methodology for information systems and software
  engineering}.
\newblock Springer, 2014.

\bibitem{winfree2017learning}
T.~Winfree, P.~Goldacre, M.~Sherkat, P.~Graham, A.~Mendoza, and T.~Miller.
\newblock Learning for low carbon living: The potential of mobile learning
  applications for built environment trades and professionals in australia.
\newblock {\em Procedia Engineering}, 180:1773--1783, 2017.

\bibitem{xu2006architectural}
L.~Xu, H.~Ziv, T.~A. Alspaugh, and D.~J. Richardson.
\newblock An architectural pattern for non-functional dependability
  requirements.
\newblock {\em Journal of Systems and Software}, 79(10):1370--1378, 2006.

\bibitem{young2003technique}
J.~Young.
\newblock {\em A technique for producing ideas}.
\newblock McGraw Hill Professional, 2003.

\bibitem{zowghi2002three}
D.~Zowghi and V.~Gervasi.
\newblock The three cs of requirements: consistency, completeness, and
  correctness.
\newblock In {\em International Workshop on Requirements Engineering:
  Foundations for Software Quality, Essen, Germany: Essener Informatik
  Beitiage}, pages 155--164, 2002.

\end{thebibliography}
